\documentclass[twocolumn,preprintnumbers,aps,showpacs,superscriptaddress,prl]{revtex4-1}

\usepackage{palatino}
\usepackage[latin1]{inputenc}
\usepackage{epsf}
\usepackage{amsmath,amssymb}
\usepackage{latexsym}
\usepackage{calc}
\usepackage{color}
\usepackage{shadow}
\usepackage{epsfig}
\usepackage{float}

\usepackage{url}

\usepackage[pdftex, pdfstartview = {FitH}]{hyperref}

\newcommand{\ben}{\begin{equation}}
\newcommand{\een}{\end{equation}}
\newcommand{\bea}{\begin{eqnarray}}
\newcommand{\eea}{\end{eqnarray}}

\def\dulr{{\underline{\underline{\bf r}}}}

\begin{document}
\title{Time-Dependent Multi-Component Density Functional Theory for Coupled Electron-Positron Dynamics}

\author{Yasumitsu Suzuki}
\affiliation{Department of Physics, Tokyo University of Science, 1-3 Kagurazaka, Shinjuku-ku, Tokyo 162-8601, Japan}  
\author{Satoshi Hagiwara}
\affiliation{Department of Physics, Tokyo University of Science, 1-3 Kagurazaka, Shinjuku-ku, Tokyo 162-8601, Japan} 
\author{Kazuyuki Watanabe}
\affiliation{Department of Physics, Tokyo University of Science, 1-3 Kagurazaka, Shinjuku-ku, Tokyo 162-8601, Japan}

\date{\today}

\begin{abstract}
Electron-positron interactions have been utilized in various fields of science.
Here we develop time-dependent multi-component density functional theory to study the
coupled electron-positron dynamics from first principles. 
We prove that there are coupled time-dependent single-particle equations
that can provide the electron and positron density dynamics,
and derive the formally exact expression for their effective potentials.
Introducing the adiabatic local density approximation to time-dependent electron-positron correlation,
we apply the theory to the dynamics of a
positronic lithium hydride molecule under a laser field. 
We demonstrate the significance of electron-positron dynamical correlation by revealing
the complex positron detachment mechanism and the suppression of electronic resonant 
excitation by the screening effect of the positron.
\end{abstract}

\maketitle

When a low-energy positron beam is directed to a material, 
the incident positron diffuses inside the bulk
 and is finally annihilated with atomic electrons
and $\gamma$ rays are emitted~\cite{positron1}.  Analysis of these $\gamma$ rays provides various information related with the surface structures, lattice defects,
and electronic structures
 of the material~\cite{positron1,positron2,positron3,*positron4}.
The analysis of positron-annihilation $\gamma$ rays
 has been utilized in many applications, such as 
positron annihilation spectroscopy~\cite{pas} and 
positron emission tomography~\cite{pet}.
The positron-material interaction plays a key role in these experiments~\cite{ps1,*ps2,*ps3}.
The interaction between positron and atom or molecule has also been widely studied to better understand
how positrons interact with atomic electrons and are bound to them~\cite{mol1,mol2,*mol3}. 
An experiment that measured positron-atom binding energies
through the study of positron-atom recombination under a laser field was recently reported~\cite{mol4}.

Theoretical approaches to study these positron physics have been
extensively developed~\cite{theo2,*theo3,*theo4,*theo5,*theo6,*theo7}.
Among them, two-component density functional theory (2C-DFT)~\cite{positron2, tcdft1,*tcdft2,*tcdft3},
which is an extension of DFT~\cite{dft1,*dft2} to the coupled electron-positron system, 
has been a powerful first-principles tool to calculate the ground-state electron and positron densities and
their properties. 
2C-DFT has been successfully applied to studies of positron interaction with atoms, solids and surfaces 
to determine the electron-positron momentum distributions~\cite{doppler}, 
positron annihilation lifetimes~\cite{surface1,*ferro1,*ferro2}, and positron binding energies~\cite{tcdftatom}, 
to name a few.
Another powerful method is the wavefunction-based approach,
such as the multi-component molecular orbital method~\cite{mcmo1,*mcmo2} and quantum
Monte Carlo method~\cite{qmc1,*qmc2,qmc3}, which have also revealed much positron physics 
with high accuracy.
However, the dynamical interaction mechanism between positrons and electrons,
especially under a laser field~\cite{mol1,mol2,*mol3,mol4}, 
 has not yet been clarified, because there has been no first-principles method
that can simulate the correlated dynamics of positrons and electrons for realistic systems.

In this study, we develop {\it time-dependent multi-component density functional theory} 
(TDMCDFT)~\cite{tdmcdft1, tdmcdft2}
 toward an understanding of the mechanism of coupled electron-positron dynamics.
Time-dependent density functional
 theory (TDDFT)~\cite{tddft1, tddft2, tddft3} has enabled us to simulate real-time many-electron dynamics, 
by mapping it to the dynamics of the non-interacting Kohn-Sham (KS) system 
evolving in a single-particle potential by virtue of the 
Runge-Gross~\cite{tddft1} and van Leeuwen~\cite{leeuwen,*leeuwen2} theorems.
TDMCDFT is an extension of TDDFT to a multi-component (MC) system,
and it has been expected to provide a first-principles simulation tool to elucidate 
the dynamics of a system that consists of different types of quantum particles.
Li and Tong proved that 
one-to-one mapping between external potentials  
and time-dependent (TD) densities is established also in MC systems~\cite{tdmcdft1}.
Gross {\it et al.} then applied the Li-Tong theorem to 
coupled electron-nuclear systems and developed TDMCKS equations for 
a one-dimensional model of the H$_2^+$ molecule~\cite{tdmcdft2}.
However, there are some difficulties to apply 
their method to real (three-dimensional) systems that consist of many electrons 
and nuclei.
To define electron and nuclear densities that are not constant in space,
a body-fixed frame transformation~\cite{mcdft1,*mcdft2}, which makes electron density reflect the internal symmetry of the system, must be conducted.  
Only a Hartree approximation has been tested for a TD electron-nuclear correlation potential, and it did not give
satisfactory results~\cite{tdmcdft2}.
It has been a challenging issue to develop electron-nuclear correlation potential functionals, 
and there have been ongoing studies~\cite{en1,*en2,*en3} including 
those based on the exact factorization approach~\cite{ef1,*ef2,*ef3,*ef4}.

Here, our purpose is to develop TDMCDFT for electron-{\it positron} dynamics.
For this, we can circumvent the problem of the body-fixed frame transformation
by treating nuclei as classical particles, because classical nuclei 
 serve as external potentials for electrons and positrons and 
the Hamiltonian of the system is no longer translationally and rotationally invariant~\footnote[1]{Note that it would also be possible to
treat nuclei quantum mechanically by developing three-component 
density functional theory. In that case the body-fixed frame transformation
would again need to be carried out
}.
 We define TD density and current density both for electrons and positrons,
and prove non-interacting $v$-representability~\cite{leeuwen,*leeuwen2}
for these quantities. 
The formally exact expression  
for the TD electron-positron correlation functional
is then derived by extending the TDDFT action principle~\cite{action} to a MC system.
Furthermore, we 
introduce the adiabatic local density approximation (ALDA)
to the TD electron-positron correlation,
and adopt the ground-state
electron-positron correlation energy functional 
within the LDA~\cite{tcdftatom}. 
Finally, the TDMCDFT method is applied to 
 the dynamics of a positronic lithium hydride ($e^+$-LiH) molecule under a laser field
and elucidate the significant role of TD electron-positron correlation in their coupled dynamics.

We begin by considering the full Hamiltonian of a system that consists of
$N^-$ electrons and $N^+$ positrons:
\ben
\hat{H}  =  \hat{T}(\dulr^-, \dulr^+) + \hat{W}(\dulr^-,\dulr^+) + \hat{V}(\dulr^-,\dulr^+, t),
\een
where the kinetic energy operator $\hat{T}=-\sum_{i=1}^{N^-} \frac{1}{2} \nabla^2_ { {\bf r}_i^- }  
-\sum_{\alpha=1}^{N^+} \frac{1}{2} \nabla^2_ { {\bf r}_\alpha^+ } $,
the interaction operator 
$\hat{W}=\sum_{i<j}^{N^-} \frac{1}{|{\bf r}_i^--{\bf r}_j^-|}
+\sum_{\alpha<\beta}^{N^+} \frac{1}{|{\bf r}_\alpha^+-{\bf r}_\beta^+|}
-\sum_{i=1}^{N^-}\sum_{\alpha=1}^{N^+} \frac{1}{|{\bf r}_i^--{\bf r}_\alpha^+|}$,
the potential operator due to the interaction with classical nuclei and TD external field
$\hat{V}=\sum_{i=1}^{N^-}v^-_{\rm ext}({\bf r}_i^-, t)+\sum_{\alpha=1}^{N^+}v^+_{\rm ext}({\bf r}_\alpha^+, t)$,
and $\dulr^-$ and $\dulr^+$ 
are the sets of 
 electronic and positronic laboratory coordinates, respectively
(i.e., $\dulr^- \equiv \{ {\bf r}^-_1,{\bf r}^-_2,\cdots,{\bf r}^-_{N^-} \}$
and $\dulr^+ \equiv \{ {\bf r}^+_1,{\bf r}^+_2,\cdots,{\bf r}^+_{N^+} \}$).
Throughout this letter, the sign $-$ ($+$) indicates an electron (positron),
and atomic units are used unless stated otherwise.
The TD electron-positron wavefunction $\Psi(\dulr^-,\dulr^+, t)$ obeys the full TD Shr{\"o}dinger
equation $i\partial_t \Psi=\hat{H}\Psi$. 
By treating nuclei as classical charges that determine the laboratory coordinate system, 
the TD electron (positron) density $n^{-(+)}$ and current density ${\bf j}^{-(+)}$ can be defined as follows:
\ben
n^\mp({\bf r}^\mp, t)  =  N^\mp \int d^{N^\pm}{\bf r}^\pm \int d^{N^\mp-1}{\bf r}^\mp |\Psi|^2,
\label{eqn:density}
\een
\ben
{\bf j}^\mp({\bf r}^\mp, t)  = \Re \left[ N^\mp \int d^{N^\pm}{\bf r}^\pm \int d^{N^\mp-1}{\bf r}^\mp 
\Psi^*(-i \nabla_{ {\bf r}^\mp }  \Psi) \right].
\een
The equations of motions for these quantities are:
\ben
\frac{\partial}{\partial t}n^\mp  =  -\nabla_{ {\bf r}^\mp}{\bf j}^\mp,
\label{eqn:eomn}
\een
\ben
\frac{\partial}{\partial t}{\bf j}^\mp  =-n^\mp(\nabla_{ {\bf r}^\mp}v_{\rm ext}^\mp)
-i \left\langle \Psi \right| [\hat{{\bf j}}^\mp, \hat{T}+\hat{W} ] \left| \Psi \right\rangle,
\label{eqn:eomj}
\een
where $\hat{{\bf j}}^\mp
=\frac{1}{2i}\sum_{l=1}^{N^\mp}\left(\nabla_{ {\bf r}_l^\mp} \delta({\bf r}^\mp-{\bf r}_l^\mp)
+\delta({\bf r}^\mp-{\bf r}_l^\mp)\nabla_{ {\bf r}_l^\mp}  \right)$.

Now we prove that, under some restrictions on the initial state, 
there are effective potentials $v_{\rm KS}^\mp$ in a non-interacting (KS) system
that reproduce
the TD electron density $n^-({\bf r}^-, t)$ and positron density $n^+({\bf r}^+, t)$
in an interacting system, i.e., TDMC non-interacting $v$-representability,
 by extending the van Leeuwen theorem~\cite{leeuwen,*leeuwen2} to the MC system.
Taking the divergence of Eqs.~(\ref{eqn:eomj}) and using the continuity
Eqs.~(\ref{eqn:eomn}) gives:
\ben
\frac{\partial^2}{\partial t^2}n^\mp  =\nabla \cdot \left(n^\mp(\nabla_{ {\bf r}^\mp}v^\mp)\right)
+i\nabla\cdot \left\langle \Psi \right| [\hat{{\bf j}}^\mp, \hat{T}+\hat{W} ] \left| \Psi \right\rangle,
\label{eqn:stdn}
\een
which is valid for both the interacting system and the KS system.
Now we impose the condition 
that the potentials in both the interacting system and the KS system are Taylor-expandable
around the initial time $t_0$, i.e.,
$v_{\rm ext/KS}^\mp({\bf r}^\mp, t)  =\sum^{\infty}_{k=0} \frac{1}{k!}v_{{\rm ext/KS},k}^\mp({\bf r}^\mp, t)(t-t_0)^k$.
Furthermore, we impose the initial conditions, i.e., that the initial state in the interacting system $\Psi_0$, and
that in KS system $\Phi_0$, yield the same densities and their first time-derivatives, i.e.,
\ben
n^\mp({\bf r}^\mp, t_0)=n_{\rm KS}^\mp({\bf r}^\mp, t_0),
\een
\ben
\left.\frac{\partial}{\partial t}n^\mp({\bf r}^\mp, t) \right|_{t=t_0}
=\left.\frac{\partial}{\partial t}n_{\rm KS}^\mp({\bf r}^\mp, t)\right|_{t=t_0}.
\een
These initial conditions uniquely determine the solutions of the
second-order differential Eqs.~(\ref{eqn:stdn}).
If $n^\mp({\bf r}^\mp, t)=n_{\rm KS}^\mp({\bf r}^\mp, t)$ at all times,
then subtracting Eqs.~(\ref{eqn:stdn}) for the interacting system and the KS system gives:
\ben
\begin{split}
\nabla &\cdot \left(n^\mp(\nabla_{ {\bf r}^\mp}(v_{\rm ext}^\mp-v_{\rm KS}^\mp ) )\right) \\
 & =i\nabla\cdot \left\langle \Phi \right| [\hat{{\bf j}}^\mp, \hat{T}] \left| \Phi \right\rangle
-i\nabla\cdot \left\langle \Psi \right| [\hat{{\bf j}}^\mp, \hat{T}+\hat{W} ] \left| \Psi \right\rangle,
\end{split}
\label{eqn:mcleeuwen1}
\een
where $\Phi$ is the full wavefunction of the KS system.
Under the boundary condition that $v_{\rm ext}^\mp-v_{\rm KS}^\mp=0$ at infinity,
 Eqs.~(\ref{eqn:mcleeuwen1}) have unique solutions for $v_{\rm KS}^\mp({\bf r}^\mp, t)$
when $n^\mp({\bf r}^\mp, t)$, $\Psi_0(\dulr^-, \dulr^+)$, $\Phi_0(\dulr^-, \dulr^+)$, 
and $v_{\rm ext}^\mp({\bf r}^\mp, t)$ are given,
similar to the procedure described in Ref.~\cite{leeuwen,*leeuwen2}.

By virtue of the TDMC non-interacting $v$-representability proved above,
the following coupled TDMCKS equations exist that produce $n^\mp({\bf r}^\mp, t)$:
\ben
i\frac{\partial}{\partial t}\psi^\mp_i({\bf r}^\mp, t)
=\left( -\frac{\nabla^2_{{\bf r}^\mp}}{2}+v_{\rm KS}^\mp({\bf r}^\mp, t) \right)\psi^\mp_i({\bf r}^\mp, t),
\label{eqn:tdmckseq}
\een
where $\sum^{N^\mp}_{l=1}\left| \psi_l^\mp({\bf r}^\mp, t)\right|^2=n^\mp({\bf r}^\mp, t)$.

Next, we derive the
expressions of $v_{\rm KS}^\mp({\bf r}^\mp, t)$ from an action principle.
Action principles in TDMCDFT can be formulated using the 
Keldysh time-contour technique~\cite{keldysh, tdmcdft2}; however, 
here we instead extend that in TDDFT formulated by Vignale~\cite{action}
to the electron-positron system.
According to the Li-Tong theorem and TDMC non-interacting $v$-representability,
the full wavefunction is a functional of the TD densities, and thus the 
quantum mechanical action 
is also a density functional:
\ben
\begin{split}
A[n^-,n^+]&=\int^{t_1}_{t_0}dt \left\langle \Psi[n^-,n^+] \right| i\partial_t-\hat{H}  \left| \Psi[n^-,n^+] \right\rangle \\
 & =A_0[n^-,n^+]-\int^{t_1}_{t_0}dt\int d{\bf r}^-n^-({\bf r}^-, t)v_{\rm ext}^-({\bf r}^-, t)\\
     &  - \int^{t_1}_{t_0}dt\int d{\bf r}^+n^+({\bf r}^+, t)v_{\rm ext}^+({\bf r}^+, t),
\end{split}
\label{eqn:action}
\een
where we define $A_0[n^-, n^+]=\int^{t_1}_{t_0}dt \left\langle \Psi \right| i\partial_t-\hat{T}-\hat{W} \left| \Psi \right\rangle$.
As in Ref.~\cite{action}, we consider the variations of the densities such that
$\delta \Psi[n^-, n^+](t_0)=0$, and the variational principle 
$\delta A[n^-, n^+]=i\left\langle \Psi(t_1) \left| \delta \Psi(t_1) \right.  \right\rangle$~\cite{tddft2,action}leads to:
 \ben
\frac{\delta A_0[n^-, n^+]}{\delta n^\mp}-v_{\rm ext}^\mp({\bf r}^\mp, t)
=i\left\langle \Psi(t_1) \left| \frac{\delta\Psi(t_1)}{\delta n^\mp}\right.  \right\rangle.
\label{eqn:variation}
\een
Similarly, the variational principle for KS systems leads to 
 \ben
\frac{\delta A_0^{\rm KS}[n^-,n^+]}{\delta n^\mp}-v_{\rm KS}^\mp({\bf r}^\mp, t)
=i\left\langle \Phi(t_1) \left| \frac{\delta\Phi(t_1)}{\delta n^\mp}\right. \right\rangle,
\label{eqn:ksvariation}
\een
where $A_0^{\rm KS}[n^-, n^+]=\int^{t_1}_{t_0}dt \left\langle \Phi \right| i\partial_t-\hat{T} \left| \Phi \right\rangle$.
Now we define
the exchange-correlation (xc) action functional for the electron-positron system as:
\ben
A^{-+}_{\rm xc}[n^-, n^+]=A_0^{\rm KS}[n^-, n^+]-A_0[n^-, n^+]-A_{\rm H}[n^-, n^+],
\label{eqn:xcaction}
\een
where $A_{\rm H}=\int dt d{\bf r}^-_1 d{\bf r}^-_2
\frac{n^-({\bf r}^-_1, t)n^-({\bf r}^-_2, t)}{2|{\bf r}^-_1 -{\bf r}^-_2|}
+\int dt d{\bf r}^- d{\bf r}^+
\frac{n^-({\bf r}^-, t)n^+({\bf r}^+, t)}{|{\bf r}^- -{\bf r}^+|}
+\int dt d{\bf r}^+_1 d{\bf r}^+_2
\frac{n^+({\bf r}^+_1, t)n^+({\bf r}^+_2, t)}{2|{\bf r}^+_1-{\bf r}^+_2| }$
is the Hartree action functional.
With these definitions of $A^{-+}_{\rm xc}$, and Eqs.~(\ref{eqn:variation}) and (\ref{eqn:ksvariation}),
we find the expression of the TDMCKS potentials:
\ben
v_{\rm KS}^\mp({\bf r}^\mp, t)=v_{\rm ext}^\mp({\bf r}^\mp, t)+v_{\rm H}^\mp({\bf r}^\mp, t)+v_{\rm xc}^\mp({\bf r}^\mp, t),
\label{eqn:tdmcksv}
\een
where $v_{\rm H}^\mp=\frac{\delta A_{\rm H}}{\delta n^\mp}$ 
are the Hartree potentials and 
\ben
v_{\rm xc}^\mp=\frac{\delta A^{-+}_{\rm xc}}{\delta n^\mp}
+i\left\langle \Psi(t_1) \left| \frac{\delta\Psi(t_1)}{\delta n^\mp}\right. \right\rangle
-i\left\langle \Phi(t_1) \left| \frac{\delta\Phi(t_1)}{\delta n^\mp}\right. \right\rangle
\label{eqn:tdvxc}
\een
are the exchange-correlation potentials, which incorporate 
all TDMC many-body effects in TDMCDFT. 
Note that these expressions~(\ref{eqn:tdvxc}) are causal because
$t_1$ can be replaced by a time infinitesimally later than $t$~\cite{tddft2,action}.

\begin{figure}[h]
 \centering
 \includegraphics*[width=1.0\columnwidth]{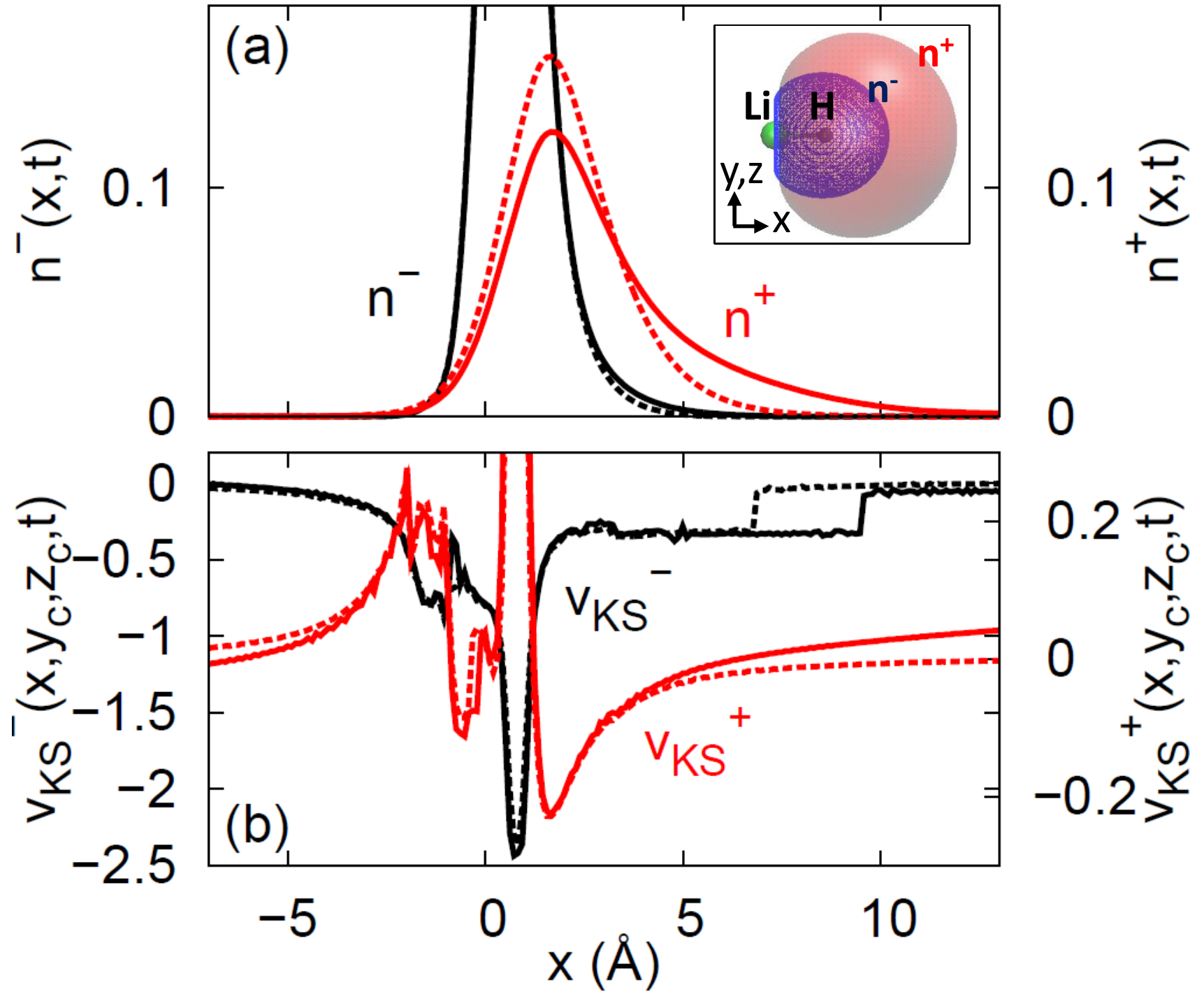}
 \caption{(a) Snapshots of $n^-(x, t)$ (black) and $n^+(x, t)$ (red) at $t=0$ (dotted) and $t=4.36$ fs (solid)
in the dynamics of $e^+$-LiH under a laser field ($\omega=1.5$ eV), and (b) corresponding $v_{\rm KS}^\mp(x,y_c,z_c, t)$. The inset shows the isosurfaces of the ground-state densities (see text).
}
 \label{fig:Fig1}
\end{figure}

Now the success of the theory is dependent on how to approximate 
$v_{\rm xc}^\mp$ so that they can be calculated practically.
One promising way to develop the approximations of $v_{\rm xc}^\mp$
will be analysis based on the exact factorization approach~\cite{ef1,*ef2,*ef3,*ef4}.
Here we introduce the adiabatic approximation,
i.e., neglecting the boundary terms and approximating $A^{-+}_{\rm xc}$ as:
\ben
A^{-+,{\rm A}}_{\rm xc}
=\int^{t_1}_{t_0}dt \left. \left( E^-_{\rm xc}[n^-_0]+E^{-+}_{\rm c}[n^-_0,n^+_0]\right)\right|_{n^\mp_0\to n^\mp(t)},
\label{eqn:alda}
\een
 where $E^-_{\rm xc}$ is the electron-electron xc energy functional in DFT, 
and $E^{-+}_{\rm c}$ is the electron-positron correlation energy functional in 2C-DFT.
Note that in Eq.~(\ref{eqn:alda}) and hereafter, we omit the positron-positron interaction term
because in many cases a system that consists of one positron and many electrons is of interest.
The adiabatic approximation to the xc term has been successfully used
in many TDDFT studies~\cite{tddft2,aldastudy1,*aldastudy2,*aldastudy3,*aldastudy4,*aldastudy5,*aldastudy6},
 while its validity and limitations have also been extensively discussed~\cite{tddft3,aldavalid1,*aldavalid2,*aldavalid3,*aldavalid4,*aldavalid5,*aldavalid6,*aldavalid7}.
Here, as the first application of TDMCDFT to a realistic molecular system, we use the adiabatic approximation.
Specifically, the LDA~\cite{lda} is used for $E^-_{\rm xc}$. For $E^{-+}_{\rm c}$, we also use the
LDA parameterized by Puska {\it et al.}, which has been reported to be suitable 
for the ground state of positronic atoms~\cite{tcdftatom}.

We now show the application of TDMCDFT presented here to
the dynamics of a $e^+$-LiH molecule under a laser field.
The LiH molecule has been the target of many previous theoretical studies on
positron-molecule interactions~\cite{mol1,qmc3,lih1,*lih2,*lih3},
and recent experimental studies~\cite{mol1,mol2,*mol3,mol4} focused on the 
response of a positron-molecule compound 
to a laser field.
Here we reveal the dynamical correlation 
between a positron and electrons and its importance
in positron detachment from LiH and electronic excitation.
One LiH molecule is placed in the center of a $30\times30\times30$ \AA$^3$ cubic unit cell
so that the molecular axis is along the $x$-axis. 
The bond length is set to the experimental value of 1.60 \AA~\cite{distance}.
One positron is then added to the LiH molecule, and the ground-state
electron and positron densities are determined by 
the self-consistent 2C-DFT calculation~\cite{positron2, tcdft1,*tcdft2,*tcdft3}
using LDA both for $E^-_{\rm xc}$~\cite{lda} and $E^{-+}_{\rm c}$~\cite{tcdftatom} 
and the plane wave basis set with norm conserving pseudopotentials~\cite{ncpp} (cut off energy 
of 816 eV). 

The inset of Fig.~\ref{fig:Fig1}(a) shows plots of the isosurfaces of 
the ground-state 
electron (blue, 0.89 e$^-/$\AA$^3$) and positron (red, 0.22 e$^+/$\AA$^3$) densities, 
where Li and H nuclei are plotted as green and gray spheres, respectively.
The electron density is localized around the H atom, and the  positron density is loosely bound to
the electron density.
The densities integrated over the $yz$-plane, 
$n^\mp(x)=\int dy \int dz n^\mp({\bf r})$, are plotted in Fig.~\ref{fig:Fig1}(a) 
as black ($n^-$) and red ($n^+$) dotted lines. The Li (H) nucleus is located at $x=-0.8$ ($0.8$) \AA.
These distributions of the ground-state densities 
and the calculated positron binding energy of 0.81 eV are in good agreement with those
calculated by the wavefunction-based approach~\cite{mol1,qmc3,lih1,*lih2,*lih3}.

\begin{figure}[h]
 \centering
\includegraphics*[width=0.9\columnwidth]{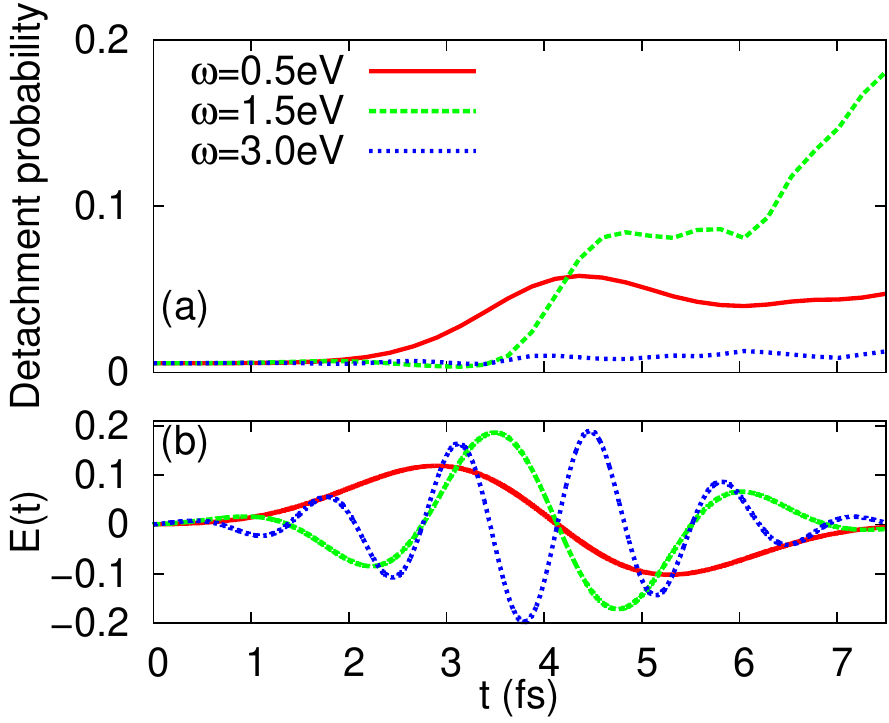}
 \caption{(a) Time evolution of positron detachment probability for three different laser fields, and
(b) corresponding $E(t)$.
}
  \label{fig:Fig2}
\end{figure}

Now we propagate these densities under the laser fields by solving the TDMCKS equations (\ref{eqn:tdmckseq})
with the adiabatic approximation to xc described above.
The laser field is applied along the $x$-axis, described within the dipole approximation
and length gauge as $v^\mp_{\rm laser}({\bf r}^\mp, t)=\mp E(t)x$
and $E(t)=E_0\sin(\omega t)\exp\left[\frac{(t-t_0)^2}{\sigma^2} \right]$,
where $E_0=0.2 $V/\AA, $\sigma=2 $fs, $t_0=4 $fs, and we compare the results from 
three different energies of $\omega=0.5$, $1.5$, and $3.0$ eV
($E(t)$ are shown in Fig.~\ref{fig:Fig2}(b)).
In Fig.~\ref{fig:Fig1}(a), the snapshot of $n^-(x, t)$ ($n^+(x, t)$) at $t=4.36$ fs for $\omega=1.5$ eV
 is plotted as a black (red) solid line~\footnote[2]{Movies of the dynamics are given in the Supplemental Material.}.
Figure~\ref{fig:Fig1}(b) shows the corresponding $v_{\rm KS}^\mp(x,y_c,z_c, t)$~\footnotemark[2]
($y_c$ ($z_c$) is the midpoint of $y$ ($z$) side of the unit cell)~\footnote[3]{Note that
steps appeared in $v_{\rm KS}^-$ around $x=5-10$ \AA are due to 
the LDA electron-positron correlation energy functional parameterized in Ref.~\cite{tcdftatom}}.
It is evident from this figure and the movies in the 
Supplemental Material~\footnotemark[2] that 
considerable positron density moves toward the positive $x$ direction at $t=4.36$ fs, which indicates
the increase of positron detachment probability.
This laser energy of $\omega=1.5$ eV gives the largest 
positron detachment probability, which is calculated by the integration of $n^+(x, t)$
over the region outside the bound region ($-3.4$ \AA $<x<7.2$ \AA )~\footnote[4]{The bound region
is defined as the region where considerable ground-state positron density exists (the region where $n^+(x, t=0)>0.001$)
},
 among the three energies as shown in Fig.~\ref{fig:Fig2}(a).
The 3.0 eV laser field leads to the lowest detachment probability.
Furthermore, the positron dynamics are not synchronized to 
$E(t)$ (Fig.~\ref{fig:Fig2}(b)), but there are retardation or even more complex response to the laser field~\footnotemark[2].
Electrons and positrons respond to the laser field in the opposite way while attracting each other.
This is the cause of the complex dynamics that appeared in Fig~\ref{fig:Fig2}.
Only the TDMCDFT calculation can reveal the mechanism as to how the coupled positron and electron respond to the laser field
by providing the dynamics of $v_{\rm KS}^\mp$. 
The movies~\footnotemark[2] show that in the case of $\omega=1.5$ eV,
 $v_{\rm KS}^+$ bends effectively for positron detachment at around $t=4$ fs,
while in the case of $\omega=3.0$ eV, $v_{\rm KS}^+$ changes the direction of its gradient
before the positron starts to depart.

\begin{figure}[h]
 \centering
 \includegraphics*[width=0.9\columnwidth]{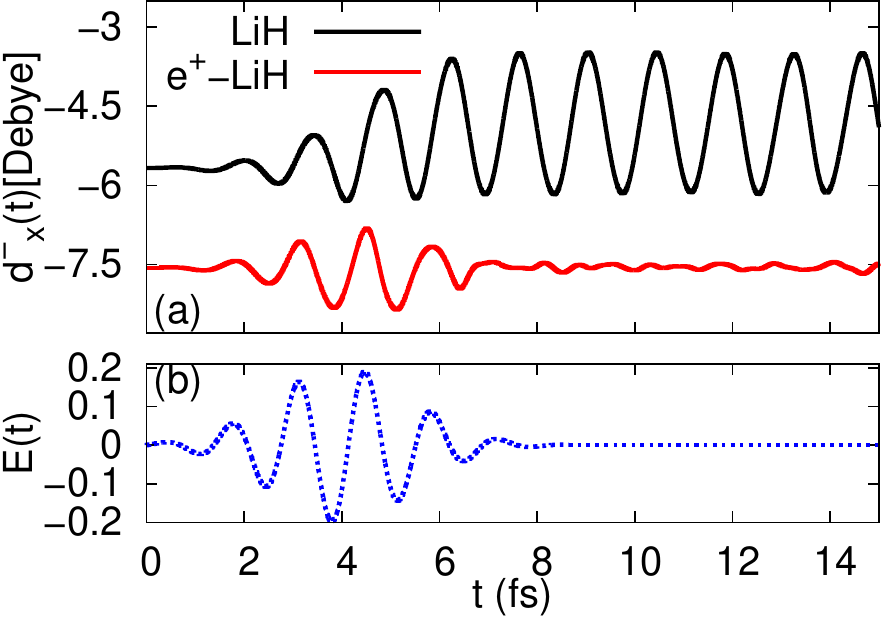}
 \caption{(a) Time-dependent
dipole moment of LiH without positronic contribution $d_x^-(t)$
of $e^+$-LiH (red) 
and LiH (black) under the laser field ($\omega=3.0$ eV), and
(b) corresponding $E(t)$.}
 \label{fig:Fig3}
\end{figure}

We present another intriguing result that shows the importance of the electron-positron dynamical correlation in Fig.~\ref{fig:Fig3}. There we compare the TD
dipole moment of LiH without positronic contribution 
$d_x^-(t)$~\footnote[5]{Time-dependent
dipole moment of LiH without positronic contribution
$d_x^-(t)$ is given by $d_x^-(t)=-\int d{\bf r}^- x n^-({\bf r}^-, t)$
}
between the system of $e^+$-LiH (red) 
and LiH (black) 
under the laser field with $\omega=3.0$ eV.
The linear-response TDDFT calculation~\cite{casida} for LiH (not shown here) 
shows that the $3.0$ eV energy laser 
elevates the electrons to 
the first excited state. 
It is evident in Fig.~\ref{fig:Fig3}
that a charge density oscillation, 
which arises from the resonant excitation and remains after the laser field fades~\cite{resonant1,*resonant2},
 is produced in LiH (black). 
The center of the dipole oscillation shifts to around $-4.85$ Debye from its ground-state value of $-5.67$ Debye, which 
indicates the occurrence of charge transfer excitation~\cite{ct1,*ct2}.
Now turning to the $e^+$-LiH case (red), we see that
such a charge oscillation is not induced, which indicates that no resonant excitation occurred.
Thus, the positron suppresses the electronic excitation,
as the result of dynamical electron-positron correlation.
A positron attracts electrons and responds to the laser field in the opposite way to electrons.
As a result, electrons are forced to move toward
 the direction opposite to the field by the positron. 
In other words, the response of the electrons to the laser field is screened by 
the positron.
The movies of the dynamics~\footnotemark[2] clearly show that
the electron density in $e^+$-LiH moves toward the opposite direction compared to that in LiH.
This mechanism of the suppression of electronic excitation by the positron can only be
found by the TDMCDFT calculation. Now it can be deduced
that positron attachment to a molecule may alter the absorption spectrum
and the excited-state nuclear dynamics.

In conclusion, we have developed
TDMCDFT for coupled electron-positron systems,
which is a first-principles method that treats both the electron and positron dynamics quantum mechanically.
TDMCDFT was applied to $e^+$-LiH
under a laser field.
There was no simple correlation between the laser energy and 
the positron detachment probability, which indicates a complex dynamical correlation between 
a positron and electrons.
Only the TDMCDFT simulation can predict the positron detachment dynamics and reveal its mechanism 
by showing the TDMCKS potentials $v_{\rm KS}^\mp$.
Furthermore, it was elucidated that
the attached positron significantly suppresses laser-induced electronic excitations,
which suggests the possibility that the absorption spectrum and excited-state nuclear dynamics may also be changed.
There are many other promising applications of the theory, such as application to the 
positron migration in the bulk of the material to reveal how it is trapped in a defect or surface,
and positron scattering by materials. These are key events in slow positron beam experiments,
and the TDMCDFT methodology developed here will be valuable for the study of fundamental positron physics.

\begin{acknowledgments}
YS and KW were supported by a Kakenhi Grants-in-Aid (No. JP16K17768 and No. JP16K05483) from the Japan Society for the Promotion of Science (JSPS).
Part of the computations were performed on
the supercomputers of the Institute for Solid State Physics,
The University of Tokyo.
\end{acknowledgments}

\bibliography{./tdmcdft}

\begin{thebibliography}{81}%
\makeatletter
\providecommand \@ifxundefined [1]{%
 \@ifx{#1\undefined}
}%
\providecommand \@ifnum [1]{%
 \ifnum #1\expandafter \@firstoftwo
 \else \expandafter \@secondoftwo
 \fi
}%
\providecommand \@ifx [1]{%
 \ifx #1\expandafter \@firstoftwo
 \else \expandafter \@secondoftwo
 \fi
}%
\providecommand \natexlab [1]{#1}%
\providecommand \enquote  [1]{``#1''}%
\providecommand \bibnamefont  [1]{#1}%
\providecommand \bibfnamefont [1]{#1}%
\providecommand \citenamefont [1]{#1}%
\providecommand \href@noop [0]{\@secondoftwo}%
\providecommand \href [0]{\begingroup \@sanitize@url \@href}%
\providecommand \@href[1]{\@@startlink{#1}\@@href}%
\providecommand \@@href[1]{\endgroup#1\@@endlink}%
\providecommand \@sanitize@url [0]{\catcode `\\12\catcode `\$12\catcode
  `\&12\catcode `\#12\catcode `\^12\catcode `\_12\catcode `\%12\relax}%
\providecommand \@@startlink[1]{}%
\providecommand \@@endlink[0]{}%
\providecommand \url  [0]{\begingroup\@sanitize@url \@url }%
\providecommand \@url [1]{\endgroup\@href {#1}{\urlprefix }}%
\providecommand \urlprefix  [0]{URL }%
\providecommand \Eprint [0]{\href }%
\providecommand \doibase [0]{http://dx.doi.org/}%
\providecommand \selectlanguage [0]{\@gobble}%
\providecommand \bibinfo  [0]{\@secondoftwo}%
\providecommand \bibfield  [0]{\@secondoftwo}%
\providecommand \translation [1]{[#1]}%
\providecommand \BibitemOpen [0]{}%
\providecommand \bibitemStop [0]{}%
\providecommand \bibitemNoStop [0]{.\EOS\space}%
\providecommand \EOS [0]{\spacefactor3000\relax}%
\providecommand \BibitemShut  [1]{\csname bibitem#1\endcsname}%
\let\auto@bib@innerbib\@empty
\bibitem [{\citenamefont {Hugenschmidt}(2016)}]{positron1}%
  \BibitemOpen
  \bibfield  {author} {\bibinfo {author} {\bibfnamefont {C.}~\bibnamefont
  {Hugenschmidt}},\ }\href@noop {} {\bibfield  {journal} {\bibinfo  {journal}
  {Surf. Sci. Rep}\ }\textbf {\bibinfo {volume} {71}},\ \bibinfo {pages} {547}
  (\bibinfo {year} {2016})}\BibitemShut {NoStop}%
\bibitem [{\citenamefont {Puska}\ and\ \citenamefont
  {Nieminen}(1994)}]{positron2}%
  \BibitemOpen
  \bibfield  {author} {\bibinfo {author} {\bibfnamefont {M.~J.}\ \bibnamefont
  {Puska}}\ and\ \bibinfo {author} {\bibfnamefont {R.~M.}\ \bibnamefont
  {Nieminen}},\ }\href@noop {} {\bibfield  {journal} {\bibinfo  {journal} {Rev.
  Mod. Phys.}\ }\textbf {\bibinfo {volume} {66}},\ \bibinfo {pages} {841}
  (\bibinfo {year} {1994})}\BibitemShut {NoStop}%
\bibitem [{\citenamefont {Eijt}\ \emph {et~al.}(2006)\citenamefont {Eijt},
  \citenamefont {van Veen}, \citenamefont {Schut}, \citenamefont {Mijnarends},
  \citenamefont {Denison}, \citenamefont {Barbiellini},\ and\ \citenamefont
  {Bansil}}]{positron3}%
  \BibitemOpen
  \bibfield  {author} {\bibinfo {author} {\bibfnamefont {S.~W.~H.}\
  \bibnamefont {Eijt}}, \bibinfo {author} {\bibfnamefont {A.}~\bibnamefont {van
  Veen}}, \bibinfo {author} {\bibfnamefont {H.}~\bibnamefont {Schut}}, \bibinfo
  {author} {\bibfnamefont {P.~E.}\ \bibnamefont {Mijnarends}}, \bibinfo
  {author} {\bibfnamefont {A.~B.}\ \bibnamefont {Denison}}, \bibinfo {author}
  {\bibfnamefont {B.}~\bibnamefont {Barbiellini}}, \ and\ \bibinfo {author}
  {\bibfnamefont {A.}~\bibnamefont {Bansil}},\ }\href@noop {} {\bibfield
  {journal} {\bibinfo  {journal} {Nat. Mater.}\ }\textbf {\bibinfo {volume}
  {5}},\ \bibinfo {pages} {23} (\bibinfo {year} {2006})}\BibitemShut {NoStop}%
\bibitem [{\citenamefont {Tuomisto}\ and\ \citenamefont
  {Makkonen}(2013)}]{positron4}%
  \BibitemOpen
  \bibfield  {author} {\bibinfo {author} {\bibfnamefont {F.}~\bibnamefont
  {Tuomisto}}\ and\ \bibinfo {author} {\bibfnamefont {I.}~\bibnamefont
  {Makkonen}},\ }\href@noop {} {\bibfield  {journal} {\bibinfo  {journal} {Rev.
  Mod. Phys.}\ }\textbf {\bibinfo {volume} {85}},\ \bibinfo {pages} {1583}
  (\bibinfo {year} {2013})}\BibitemShut {NoStop}%
\bibitem [{\citenamefont {Gidley}\ \emph {et~al.}(2006)\citenamefont {Gidley},
  \citenamefont {Peng},\ and\ \citenamefont {Vallery}}]{pas}%
  \BibitemOpen
  \bibfield  {author} {\bibinfo {author} {\bibfnamefont {D.~W.}\ \bibnamefont
  {Gidley}}, \bibinfo {author} {\bibfnamefont {H.-G.}\ \bibnamefont {Peng}}, \
  and\ \bibinfo {author} {\bibfnamefont {R.~S.}\ \bibnamefont {Vallery}},\
  }\href@noop {} {\bibfield  {journal} {\bibinfo  {journal} {Ann. Rev. Mater.
  Res.}\ }\textbf {\bibinfo {volume} {36}},\ \bibinfo {pages} {49} (\bibinfo
  {year} {2006})}\BibitemShut {NoStop}%
\bibitem [{\citenamefont {Wahl}(2002)}]{pet}%
  \BibitemOpen
  \bibfield  {author} {\bibinfo {author} {\bibfnamefont {R.~L.}\ \bibnamefont
  {Wahl}},\ }\href@noop {} {\emph {\bibinfo {title} {Principles and Practice of
  Positron Emission Tomography}}}\ (\bibinfo  {publisher} {Lippincott, Williams
  and Wilkins, Phyladelphia},\ \bibinfo {year} {2002})\BibitemShut {NoStop}%
\bibitem [{\citenamefont {Jones}\ \emph {et~al.}(2016)\citenamefont {Jones},
  \citenamefont {Rutbeck-Goldman}, \citenamefont {Hisakado}, \citenamefont
  {Pi{\~n}eiro}, \citenamefont {Tom}, \citenamefont {Mills}, \citenamefont
  {Barbiellini},\ and\ \citenamefont {Kuriplach}}]{ps1}%
  \BibitemOpen
  \bibfield  {author} {\bibinfo {author} {\bibfnamefont {A.~C.~L.}\
  \bibnamefont {Jones}}, \bibinfo {author} {\bibfnamefont {H.~J.}\ \bibnamefont
  {Rutbeck-Goldman}}, \bibinfo {author} {\bibfnamefont {T.~H.}\ \bibnamefont
  {Hisakado}}, \bibinfo {author} {\bibfnamefont {A.~M.}\ \bibnamefont
  {Pi{\~n}eiro}}, \bibinfo {author} {\bibfnamefont {H.~W.~K.}\ \bibnamefont
  {Tom}}, \bibinfo {author} {\bibfnamefont {A.~P.}\ \bibnamefont {Mills},
  \bibfnamefont {Jr.}}, \bibinfo {author} {\bibfnamefont {B.}~\bibnamefont
  {Barbiellini}}, \ and\ \bibinfo {author} {\bibfnamefont {J.}~\bibnamefont
  {Kuriplach}},\ }\href@noop {} {\bibfield  {journal} {\bibinfo  {journal}
  {Phys. Rev. Lett.}\ }\textbf {\bibinfo {volume} {117}},\ \bibinfo {pages}
  {216402} (\bibinfo {year} {2016})}\BibitemShut {NoStop}%
\bibitem [{\citenamefont {Tachibana}\ \emph {et~al.}(2018)\citenamefont
  {Tachibana}, \citenamefont {Yamashita}, \citenamefont {Nagira}, \citenamefont
  {Yabuki},\ and\ \citenamefont {Nagashima}}]{ps2}%
  \BibitemOpen
  \bibfield  {author} {\bibinfo {author} {\bibfnamefont {T.}~\bibnamefont
  {Tachibana}}, \bibinfo {author} {\bibfnamefont {T.}~\bibnamefont
  {Yamashita}}, \bibinfo {author} {\bibfnamefont {M.}~\bibnamefont {Nagira}},
  \bibinfo {author} {\bibfnamefont {H.}~\bibnamefont {Yabuki}}, \ and\ \bibinfo
  {author} {\bibfnamefont {Y.}~\bibnamefont {Nagashima}},\ }\href@noop {}
  {\bibfield  {journal} {\bibinfo  {journal} {Sci. Rep.}\ }\textbf {\bibinfo
  {volume} {8}},\ \bibinfo {pages} {7197} (\bibinfo {year} {2018})}\BibitemShut
  {NoStop}%
\bibitem [{\citenamefont {Michishio}\ \emph {et~al.}(2011)\citenamefont
  {Michishio}, \citenamefont {Tachibana}, \citenamefont {Terabe}, \citenamefont
  {Igarashi}, \citenamefont {Wada}, \citenamefont {Kuga}, \citenamefont
  {Yagishita}, \citenamefont {Hyodo},\ and\ \citenamefont {Nagashima}}]{ps3}%
  \BibitemOpen
  \bibfield  {author} {\bibinfo {author} {\bibfnamefont {K.}~\bibnamefont
  {Michishio}}, \bibinfo {author} {\bibfnamefont {T.}~\bibnamefont
  {Tachibana}}, \bibinfo {author} {\bibfnamefont {H.}~\bibnamefont {Terabe}},
  \bibinfo {author} {\bibfnamefont {A.}~\bibnamefont {Igarashi}}, \bibinfo
  {author} {\bibfnamefont {K.}~\bibnamefont {Wada}}, \bibinfo {author}
  {\bibfnamefont {T.}~\bibnamefont {Kuga}}, \bibinfo {author} {\bibfnamefont
  {A.}~\bibnamefont {Yagishita}}, \bibinfo {author} {\bibfnamefont
  {T.}~\bibnamefont {Hyodo}}, \ and\ \bibinfo {author} {\bibfnamefont
  {Y.}~\bibnamefont {Nagashima}},\ }\href@noop {} {\bibfield  {journal}
  {\bibinfo  {journal} {Phys. Rev. Lett.}\ }\textbf {\bibinfo {volume} {106}},\
  \bibinfo {pages} {153401} (\bibinfo {year} {2011})}\BibitemShut {NoStop}%
\bibitem [{\citenamefont {Gribakin}\ \emph {et~al.}(2010)\citenamefont
  {Gribakin}, \citenamefont {Young},\ and\ \citenamefont {Surko}}]{mol1}%
  \BibitemOpen
  \bibfield  {author} {\bibinfo {author} {\bibfnamefont {G.~F.}\ \bibnamefont
  {Gribakin}}, \bibinfo {author} {\bibfnamefont {J.~A.}\ \bibnamefont {Young}},
  \ and\ \bibinfo {author} {\bibfnamefont {C.~M.}\ \bibnamefont {Surko}},\
  }\href@noop {} {\bibfield  {journal} {\bibinfo  {journal} {Rev. Mod. Phys.}\
  }\textbf {\bibinfo {volume} {82}},\ \bibinfo {pages} {2557} (\bibinfo {year}
  {2010})}\BibitemShut {NoStop}%
\bibitem [{\citenamefont {Danielson}\ \emph {et~al.}(2012)\citenamefont
  {Danielson}, \citenamefont {Jones}, \citenamefont {Natisin},\ and\
  \citenamefont {Surko}}]{mol2}%
  \BibitemOpen
  \bibfield  {author} {\bibinfo {author} {\bibfnamefont {J.~R.}\ \bibnamefont
  {Danielson}}, \bibinfo {author} {\bibfnamefont {A.~C.~L.}\ \bibnamefont
  {Jones}}, \bibinfo {author} {\bibfnamefont {M.~R.}\ \bibnamefont {Natisin}},
  \ and\ \bibinfo {author} {\bibfnamefont {C.~M.}\ \bibnamefont {Surko}},\
  }\href@noop {} {\bibfield  {journal} {\bibinfo  {journal} {Phys. Rev. Lett.}\
  }\textbf {\bibinfo {volume} {109}},\ \bibinfo {pages} {113201} (\bibinfo
  {year} {2012})}\BibitemShut {NoStop}%
\bibitem [{\citenamefont {Danielson}\ \emph {et~al.}(2010)\citenamefont
  {Danielson}, \citenamefont {Gosselin},\ and\ \citenamefont {Surko}}]{mol3}%
  \BibitemOpen
  \bibfield  {author} {\bibinfo {author} {\bibfnamefont {J.~R.}\ \bibnamefont
  {Danielson}}, \bibinfo {author} {\bibfnamefont {J.~J.}\ \bibnamefont
  {Gosselin}}, \ and\ \bibinfo {author} {\bibfnamefont {C.~M.}\ \bibnamefont
  {Surko}},\ }\href@noop {} {\bibfield  {journal} {\bibinfo  {journal} {Phys.
  Rev. Lett.}\ }\textbf {\bibinfo {volume} {104}},\ \bibinfo {pages} {233201}
  (\bibinfo {year} {2010})}\BibitemShut {NoStop}%
\bibitem [{\citenamefont {Surko}\ \emph {et~al.}(2012)\citenamefont {Surko},
  \citenamefont {Danielson}, \citenamefont {Gribakin},\ and\ \citenamefont
  {Continetti}}]{mol4}%
  \BibitemOpen
  \bibfield  {author} {\bibinfo {author} {\bibfnamefont {C.~M.}\ \bibnamefont
  {Surko}}, \bibinfo {author} {\bibfnamefont {J.~R.}\ \bibnamefont
  {Danielson}}, \bibinfo {author} {\bibfnamefont {G.~F.}\ \bibnamefont
  {Gribakin}}, \ and\ \bibinfo {author} {\bibfnamefont {R.~E.}\ \bibnamefont
  {Continetti}},\ }\href@noop {} {\bibfield  {journal} {\bibinfo  {journal}
  {New J. Phys.}\ }\textbf {\bibinfo {volume} {14}},\ \bibinfo {pages} {065004}
  (\bibinfo {year} {2012})}\BibitemShut {NoStop}%
\bibitem [{\citenamefont {Mitroy}\ \emph {et~al.}(2002)\citenamefont {Mitroy},
  \citenamefont {Bromley},\ and\ \citenamefont {Ryzhikh}}]{theo2}%
  \BibitemOpen
  \bibfield  {author} {\bibinfo {author} {\bibfnamefont {J.}~\bibnamefont
  {Mitroy}}, \bibinfo {author} {\bibfnamefont {M.~W.~J.}\ \bibnamefont
  {Bromley}}, \ and\ \bibinfo {author} {\bibfnamefont {G.~G.}\ \bibnamefont
  {Ryzhikh}},\ }\href@noop {} {\bibfield  {journal} {\bibinfo  {journal} {J.
  Phys. B}\ }\textbf {\bibinfo {volume} {35}},\ \bibinfo {pages} {R81}
  (\bibinfo {year} {2002})}\BibitemShut {NoStop}%
\bibitem [{\citenamefont {Green}(2017)}]{theo3}%
  \BibitemOpen
  \bibfield  {author} {\bibinfo {author} {\bibfnamefont {D.~G.}\ \bibnamefont
  {Green}},\ }\href@noop {} {\bibfield  {journal} {\bibinfo  {journal} {Phys.
  Rev. Lett.}\ }\textbf {\bibinfo {volume} {119}},\ \bibinfo {pages} {203404}
  (\bibinfo {year} {2017})}\BibitemShut {NoStop}%
\bibitem [{\citenamefont {Bubin}\ and\ \citenamefont
  {Adamowicz}(2006)}]{theo4}%
  \BibitemOpen
  \bibfield  {author} {\bibinfo {author} {\bibfnamefont {S.}~\bibnamefont
  {Bubin}}\ and\ \bibinfo {author} {\bibfnamefont {L.}~\bibnamefont
  {Adamowicz}},\ }\href@noop {} {\bibfield  {journal} {\bibinfo  {journal}
  {Phys. Rev. A}\ }\textbf {\bibinfo {volume} {74}},\ \bibinfo {pages} {052502}
  (\bibinfo {year} {2006})}\BibitemShut {NoStop}%
\bibitem [{\citenamefont {Pak}\ \emph {et~al.}(2009)\citenamefont {Pak},
  \citenamefont {Chakraborty},\ and\ \citenamefont {Hammes-Schiffer}}]{theo5}%
  \BibitemOpen
  \bibfield  {author} {\bibinfo {author} {\bibfnamefont {M.~V.}\ \bibnamefont
  {Pak}}, \bibinfo {author} {\bibfnamefont {A.}~\bibnamefont {Chakraborty}}, \
  and\ \bibinfo {author} {\bibfnamefont {S.}~\bibnamefont {Hammes-Schiffer}},\
  }\href@noop {} {\bibfield  {journal} {\bibinfo  {journal} {J. Phys. Chem. A}\
  }\textbf {\bibinfo {volume} {113}},\ \bibinfo {pages} {4004} (\bibinfo {year}
  {2009})}\BibitemShut {NoStop}%
\bibitem [{\citenamefont {Usukura}\ \emph {et~al.}(1998)\citenamefont
  {Usukura}, \citenamefont {Varga},\ and\ \citenamefont {Suzuki}}]{theo6}%
  \BibitemOpen
  \bibfield  {author} {\bibinfo {author} {\bibfnamefont {J.}~\bibnamefont
  {Usukura}}, \bibinfo {author} {\bibfnamefont {K.}~\bibnamefont {Varga}}, \
  and\ \bibinfo {author} {\bibfnamefont {Y.}~\bibnamefont {Suzuki}},\
  }\href@noop {} {\bibfield  {journal} {\bibinfo  {journal} {Phys. Rev. A}\
  }\textbf {\bibinfo {volume} {58}},\ \bibinfo {pages} {1918} (\bibinfo {year}
  {1998})}\BibitemShut {NoStop}%
\bibitem [{\citenamefont {Zubiaga}\ \emph
  {et~al.}(2014{\natexlab{a}})\citenamefont {Zubiaga}, \citenamefont
  {Tuomisto},\ and\ \citenamefont {Puska}}]{theo7}%
  \BibitemOpen
  \bibfield  {author} {\bibinfo {author} {\bibfnamefont {A.}~\bibnamefont
  {Zubiaga}}, \bibinfo {author} {\bibfnamefont {F.}~\bibnamefont {Tuomisto}}, \
  and\ \bibinfo {author} {\bibfnamefont {M.~J.}\ \bibnamefont {Puska}},\
  }\href@noop {} {\bibfield  {journal} {\bibinfo  {journal} {J. Phys. Chem. B}\
  }\textbf {\bibinfo {volume} {119}},\ \bibinfo {pages} {1747} (\bibinfo {year}
  {2014}{\natexlab{a}})}\BibitemShut {NoStop}%
\bibitem [{\citenamefont {Boro{\'n}ski}\ and\ \citenamefont
  {Nieminen}(1986)}]{tcdft1}%
  \BibitemOpen
  \bibfield  {author} {\bibinfo {author} {\bibfnamefont {E.}~\bibnamefont
  {Boro{\'n}ski}}\ and\ \bibinfo {author} {\bibfnamefont {R.~M.}\ \bibnamefont
  {Nieminen}},\ }\href@noop {} {\bibfield  {journal} {\bibinfo  {journal}
  {Phys. Rev. B}\ }\textbf {\bibinfo {volume} {34}},\ \bibinfo {pages} {3820}
  (\bibinfo {year} {1986})}\BibitemShut {NoStop}%
\bibitem [{\citenamefont {Puska}\ \emph {et~al.}(1995)\citenamefont {Puska},
  \citenamefont {Seitsonen},\ and\ \citenamefont {Nieminen}}]{tcdft2}%
  \BibitemOpen
  \bibfield  {author} {\bibinfo {author} {\bibfnamefont {M.~J.}\ \bibnamefont
  {Puska}}, \bibinfo {author} {\bibfnamefont {A.~P.}\ \bibnamefont
  {Seitsonen}}, \ and\ \bibinfo {author} {\bibfnamefont {R.~M.}\ \bibnamefont
  {Nieminen}},\ }\href@noop {} {\bibfield  {journal} {\bibinfo  {journal}
  {Phys. Rev. B}\ }\textbf {\bibinfo {volume} {52}},\ \bibinfo {pages} {10947}
  (\bibinfo {year} {1995})}\BibitemShut {NoStop}%
\bibitem [{\citenamefont {Barbiellini}\ and\ \citenamefont
  {Kuriplach}(2015)}]{tcdft3}%
  \BibitemOpen
  \bibfield  {author} {\bibinfo {author} {\bibfnamefont {B.}~\bibnamefont
  {Barbiellini}}\ and\ \bibinfo {author} {\bibfnamefont {J.}~\bibnamefont
  {Kuriplach}},\ }\href@noop {} {\bibfield  {journal} {\bibinfo  {journal}
  {Phys. Rev. Lett.}\ }\textbf {\bibinfo {volume} {114}},\ \bibinfo {pages}
  {147401} (\bibinfo {year} {2015})}\BibitemShut {NoStop}%
\bibitem [{\citenamefont {Hohenberg}\ and\ \citenamefont {Kohn}(1964)}]{dft1}%
  \BibitemOpen
  \bibfield  {author} {\bibinfo {author} {\bibfnamefont {P.}~\bibnamefont
  {Hohenberg}}\ and\ \bibinfo {author} {\bibfnamefont {W.}~\bibnamefont
  {Kohn}},\ }\href@noop {} {\bibfield  {journal} {\bibinfo  {journal} {Phys.
  Rev}\ }\textbf {\bibinfo {volume} {136}},\ \bibinfo {pages} {B864} (\bibinfo
  {year} {1964})}\BibitemShut {NoStop}%
\bibitem [{\citenamefont {Kohn}\ and\ \citenamefont {Sham}(1965)}]{dft2}%
  \BibitemOpen
  \bibfield  {author} {\bibinfo {author} {\bibfnamefont {W.}~\bibnamefont
  {Kohn}}\ and\ \bibinfo {author} {\bibfnamefont {L.~J.}\ \bibnamefont
  {Sham}},\ }\href@noop {} {\bibfield  {journal} {\bibinfo  {journal} {Phys.
  Rev}\ }\textbf {\bibinfo {volume} {140}},\ \bibinfo {pages} {A1133} (\bibinfo
  {year} {1965})}\BibitemShut {NoStop}%
\bibitem [{\citenamefont {Uedono}\ \emph {et~al.}(2012)\citenamefont {Uedono},
  \citenamefont {Ishibashi}, \citenamefont {Tenjinbayashi}, \citenamefont
  {Tsutsui}, \citenamefont {Nakahara}, \citenamefont {Takamizu},\ and\
  \citenamefont {Chichibu}}]{doppler}%
  \BibitemOpen
  \bibfield  {author} {\bibinfo {author} {\bibfnamefont {A.}~\bibnamefont
  {Uedono}}, \bibinfo {author} {\bibfnamefont {S.}~\bibnamefont {Ishibashi}},
  \bibinfo {author} {\bibfnamefont {K.}~\bibnamefont {Tenjinbayashi}}, \bibinfo
  {author} {\bibfnamefont {T.}~\bibnamefont {Tsutsui}}, \bibinfo {author}
  {\bibfnamefont {K.}~\bibnamefont {Nakahara}}, \bibinfo {author}
  {\bibfnamefont {D.}~\bibnamefont {Takamizu}}, \ and\ \bibinfo {author}
  {\bibfnamefont {S.~F.}\ \bibnamefont {Chichibu}},\ }\href@noop {} {\bibfield
  {journal} {\bibinfo  {journal} {J. Appl. Phys.}\ }\textbf {\bibinfo {volume}
  {111}},\ \bibinfo {pages} {014508} (\bibinfo {year} {2012})}\BibitemShut
  {NoStop}%
\bibitem [{\citenamefont {Hagiwara}\ \emph {et~al.}(2015)\citenamefont
  {Hagiwara}, \citenamefont {Hu},\ and\ \citenamefont {Watanabe}}]{surface1}%
  \BibitemOpen
  \bibfield  {author} {\bibinfo {author} {\bibfnamefont {S.}~\bibnamefont
  {Hagiwara}}, \bibinfo {author} {\bibfnamefont {C.}~\bibnamefont {Hu}}, \ and\
  \bibinfo {author} {\bibfnamefont {K.}~\bibnamefont {Watanabe}},\ }\href@noop
  {} {\bibfield  {journal} {\bibinfo  {journal} {Phys. Rev. B}\ }\textbf
  {\bibinfo {volume} {91}},\ \bibinfo {pages} {115409} (\bibinfo {year}
  {2015})}\BibitemShut {NoStop}%
\bibitem [{\citenamefont {Lin}\ \emph {et~al.}(2014)\citenamefont {Lin},
  \citenamefont {Yamasaki},\ and\ \citenamefont {Saito}}]{ferro1}%
  \BibitemOpen
  \bibfield  {author} {\bibinfo {author} {\bibfnamefont {J.}~\bibnamefont
  {Lin}}, \bibinfo {author} {\bibfnamefont {T.}~\bibnamefont {Yamasaki}}, \
  and\ \bibinfo {author} {\bibfnamefont {M.}~\bibnamefont {Saito}},\
  }\href@noop {} {\bibfield  {journal} {\bibinfo  {journal} {Jpn. J. Appl.
  Phys.}\ }\textbf {\bibinfo {volume} {53}},\ \bibinfo {pages} {053002}
  (\bibinfo {year} {2014})}\BibitemShut {NoStop}%
\bibitem [{\citenamefont {Hagiwara}\ \emph {et~al.}(2017)\citenamefont
  {Hagiwara}, \citenamefont {Suzuki},\ and\ \citenamefont {Watanabe}}]{ferro2}%
  \BibitemOpen
  \bibfield  {author} {\bibinfo {author} {\bibfnamefont {S.}~\bibnamefont
  {Hagiwara}}, \bibinfo {author} {\bibfnamefont {Y.}~\bibnamefont {Suzuki}}, \
  and\ \bibinfo {author} {\bibfnamefont {K.}~\bibnamefont {Watanabe}},\
  }\href@noop {} {\bibfield  {journal} {\bibinfo  {journal} {Appl. Phys.
  Express}\ }\textbf {\bibinfo {volume} {10}},\ \bibinfo {pages} {045101}
  (\bibinfo {year} {2017})}\BibitemShut {NoStop}%
\bibitem [{\citenamefont {Zubiaga}\ \emph
  {et~al.}(2014{\natexlab{b}})\citenamefont {Zubiaga}, \citenamefont
  {Tuomisto},\ and\ \citenamefont {Puska}}]{tcdftatom}%
  \BibitemOpen
  \bibfield  {author} {\bibinfo {author} {\bibfnamefont {A.}~\bibnamefont
  {Zubiaga}}, \bibinfo {author} {\bibfnamefont {F.}~\bibnamefont {Tuomisto}}, \
  and\ \bibinfo {author} {\bibfnamefont {M.~J.}\ \bibnamefont {Puska}},\
  }\href@noop {} {\bibfield  {journal} {\bibinfo  {journal} {Phys. Rev. A}\
  }\textbf {\bibinfo {volume} {89}},\ \bibinfo {pages} {052707} (\bibinfo
  {year} {2014}{\natexlab{b}})}\BibitemShut {NoStop}%
\bibitem [{\citenamefont {Kurtz}\ and\ \citenamefont {Jordan}(1981)}]{mcmo1}%
  \BibitemOpen
  \bibfield  {author} {\bibinfo {author} {\bibfnamefont {H.~A.}\ \bibnamefont
  {Kurtz}}\ and\ \bibinfo {author} {\bibfnamefont {K.~D.}\ \bibnamefont
  {Jordan}},\ }\href@noop {} {\bibfield  {journal} {\bibinfo  {journal} {J.
  Chem. Phys.}\ }\textbf {\bibinfo {volume} {75}},\ \bibinfo {pages} {1876}
  (\bibinfo {year} {1981})}\BibitemShut {NoStop}%
\bibitem [{\citenamefont {Tachikawa}\ \emph {et~al.}(2011)\citenamefont
  {Tachikawa}, \citenamefont {Kita},\ and\ \citenamefont {Buenker}}]{mcmo2}%
  \BibitemOpen
  \bibfield  {author} {\bibinfo {author} {\bibfnamefont {M.}~\bibnamefont
  {Tachikawa}}, \bibinfo {author} {\bibfnamefont {Y.}~\bibnamefont {Kita}}, \
  and\ \bibinfo {author} {\bibfnamefont {R.~J.}\ \bibnamefont {Buenker}},\
  }\href@noop {} {\bibfield  {journal} {\bibinfo  {journal} {Phys. Chem. Chem.
  Phys.}\ }\textbf {\bibinfo {volume} {13}},\ \bibinfo {pages} {2701} (\bibinfo
  {year} {2011})}\BibitemShut {NoStop}%
\bibitem [{\citenamefont {Schrader}\ \emph {et~al.}(1992)\citenamefont
  {Schrader}, \citenamefont {Yoshida},\ and\ \citenamefont {Iguchi}}]{qmc1}%
  \BibitemOpen
  \bibfield  {author} {\bibinfo {author} {\bibfnamefont {D.~M.}\ \bibnamefont
  {Schrader}}, \bibinfo {author} {\bibfnamefont {T.}~\bibnamefont {Yoshida}}, \
  and\ \bibinfo {author} {\bibfnamefont {K.}~\bibnamefont {Iguchi}},\
  }\href@noop {} {\bibfield  {journal} {\bibinfo  {journal} {Phys. Rev. Lett.}\
  }\textbf {\bibinfo {volume} {68}},\ \bibinfo {pages} {3281} (\bibinfo {year}
  {1992})}\BibitemShut {NoStop}%
\bibitem [{\citenamefont {Bressanini}\ \emph {et~al.}(1998)\citenamefont
  {Bressanini}, \citenamefont {Mella},\ and\ \citenamefont {Morosi}}]{qmc2}%
  \BibitemOpen
  \bibfield  {author} {\bibinfo {author} {\bibfnamefont {D.}~\bibnamefont
  {Bressanini}}, \bibinfo {author} {\bibfnamefont {M.}~\bibnamefont {Mella}}, \
  and\ \bibinfo {author} {\bibfnamefont {G.}~\bibnamefont {Morosi}},\
  }\href@noop {} {\bibfield  {journal} {\bibinfo  {journal} {J. Chem. Phys.}\
  }\textbf {\bibinfo {volume} {108}},\ \bibinfo {pages} {4756} (\bibinfo {year}
  {1998})}\BibitemShut {NoStop}%
\bibitem [{\citenamefont {Kita}\ \emph {et~al.}(2009)\citenamefont {Kita},
  \citenamefont {Maezono}, \citenamefont {Tachikawa}, \citenamefont {Towler},\
  and\ \citenamefont {Needs}}]{qmc3}%
  \BibitemOpen
  \bibfield  {author} {\bibinfo {author} {\bibfnamefont {Y.}~\bibnamefont
  {Kita}}, \bibinfo {author} {\bibfnamefont {R.}~\bibnamefont {Maezono}},
  \bibinfo {author} {\bibfnamefont {M.}~\bibnamefont {Tachikawa}}, \bibinfo
  {author} {\bibfnamefont {M.}~\bibnamefont {Towler}}, \ and\ \bibinfo {author}
  {\bibfnamefont {R.~J.}\ \bibnamefont {Needs}},\ }\href@noop {} {\bibfield
  {journal} {\bibinfo  {journal} {J. Chem. Phys.}\ }\textbf {\bibinfo {volume}
  {131}},\ \bibinfo {pages} {134310} (\bibinfo {year} {2009})}\BibitemShut
  {NoStop}%
\bibitem [{\citenamefont {Li}\ and\ \citenamefont {Tong}(1986)}]{tdmcdft1}%
  \BibitemOpen
  \bibfield  {author} {\bibinfo {author} {\bibfnamefont {T.-C.}\ \bibnamefont
  {Li}}\ and\ \bibinfo {author} {\bibfnamefont {P.-Q.}\ \bibnamefont {Tong}},\
  }\href@noop {} {\bibfield  {journal} {\bibinfo  {journal} {Phys. Rev. A}\
  }\textbf {\bibinfo {volume} {34}},\ \bibinfo {pages} {529} (\bibinfo {year}
  {1986})}\BibitemShut {NoStop}%
\bibitem [{\citenamefont {Butriy}\ \emph {et~al.}(2007)\citenamefont {Butriy},
  \citenamefont {Ebadi}, \citenamefont {de~Boeij}, \citenamefont {van
  Leeuwen},\ and\ \citenamefont {Gross}}]{tdmcdft2}%
  \BibitemOpen
  \bibfield  {author} {\bibinfo {author} {\bibfnamefont {O.}~\bibnamefont
  {Butriy}}, \bibinfo {author} {\bibfnamefont {H.}~\bibnamefont {Ebadi}},
  \bibinfo {author} {\bibfnamefont {P.~L.}\ \bibnamefont {de~Boeij}}, \bibinfo
  {author} {\bibfnamefont {R.}~\bibnamefont {van Leeuwen}}, \ and\ \bibinfo
  {author} {\bibfnamefont {E.~K.~U.}\ \bibnamefont {Gross}},\ }\href@noop {}
  {\bibfield  {journal} {\bibinfo  {journal} {Phys. Rev. A}\ }\textbf {\bibinfo
  {volume} {76}},\ \bibinfo {pages} {052514} (\bibinfo {year}
  {2007})}\BibitemShut {NoStop}%
\bibitem [{\citenamefont {Runge}\ and\ \citenamefont {Gross}(1984)}]{tddft1}%
  \BibitemOpen
  \bibfield  {author} {\bibinfo {author} {\bibfnamefont {E.}~\bibnamefont
  {Runge}}\ and\ \bibinfo {author} {\bibfnamefont {E.~K.~U.}\ \bibnamefont
  {Gross}},\ }\href@noop {} {\bibfield  {journal} {\bibinfo  {journal} {Phys.
  Rev. Lett.}\ }\textbf {\bibinfo {volume} {52}},\ \bibinfo {pages} {997}
  (\bibinfo {year} {1984})}\BibitemShut {NoStop}%
\bibitem [{\citenamefont {Ullrich}(2012)}]{tddft2}%
  \BibitemOpen
  \bibfield  {author} {\bibinfo {author} {\bibfnamefont {C.~A.}\ \bibnamefont
  {Ullrich}},\ }\href@noop {} {\emph {\bibinfo {title} {Time-Dependent
  Density-Functional Theory: Concepts and Applications}}}\ (\bibinfo
  {publisher} {Oxford University Press},\ \bibinfo {year} {2012})\BibitemShut
  {NoStop}%
\bibitem [{\citenamefont {Maitra}(2016)}]{tddft3}%
  \BibitemOpen
  \bibfield  {author} {\bibinfo {author} {\bibfnamefont {N.~T.}\ \bibnamefont
  {Maitra}},\ }\href@noop {} {\bibfield  {journal} {\bibinfo  {journal} {J.
  Chem. Phys.}\ }\textbf {\bibinfo {volume} {144}},\ \bibinfo {pages} {220901}
  (\bibinfo {year} {2016})}\BibitemShut {NoStop}%
\bibitem [{\citenamefont {van Leeuwen}(1999)}]{leeuwen}%
  \BibitemOpen
  \bibfield  {author} {\bibinfo {author} {\bibfnamefont {R.}~\bibnamefont {van
  Leeuwen}},\ }\href@noop {} {\bibfield  {journal} {\bibinfo  {journal} {Phys.
  Rev. Lett.}\ }\textbf {\bibinfo {volume} {82}},\ \bibinfo {pages} {3863}
  (\bibinfo {year} {1999})}\BibitemShut {NoStop}%
\bibitem [{\citenamefont {Ruggenthaler}\ \emph {et~al.}(2015)\citenamefont
  {Ruggenthaler}, \citenamefont {Penz},\ and\ \citenamefont {van
  Leeuwen}}]{leeuwen2}%
  \BibitemOpen
  \bibfield  {author} {\bibinfo {author} {\bibfnamefont {M.}~\bibnamefont
  {Ruggenthaler}}, \bibinfo {author} {\bibfnamefont {M.}~\bibnamefont {Penz}},
  \ and\ \bibinfo {author} {\bibfnamefont {R.}~\bibnamefont {van Leeuwen}},\
  }\href@noop {} {\bibfield  {journal} {\bibinfo  {journal} {J. Phys. Condens.
  Matter}\ }\textbf {\bibinfo {volume} {27}},\ \bibinfo {pages} {203202}
  (\bibinfo {year} {2015})}\BibitemShut {NoStop}%
\bibitem [{\citenamefont {Kreibich}\ and\ \citenamefont
  {Gross}(2001)}]{mcdft1}%
  \BibitemOpen
  \bibfield  {author} {\bibinfo {author} {\bibfnamefont {T.}~\bibnamefont
  {Kreibich}}\ and\ \bibinfo {author} {\bibfnamefont {E.~K.~U.}\ \bibnamefont
  {Gross}},\ }\href@noop {} {\bibfield  {journal} {\bibinfo  {journal} {Phys.
  Rev. Lett.}\ }\textbf {\bibinfo {volume} {86}},\ \bibinfo {pages} {2984}
  (\bibinfo {year} {2001})}\BibitemShut {NoStop}%
\bibitem [{\citenamefont {Kreibich}\ \emph {et~al.}(2008)\citenamefont
  {Kreibich}, \citenamefont {van Leeuwen},\ and\ \citenamefont
  {Gross}}]{mcdft2}%
  \BibitemOpen
  \bibfield  {author} {\bibinfo {author} {\bibfnamefont {T.}~\bibnamefont
  {Kreibich}}, \bibinfo {author} {\bibfnamefont {R.}~\bibnamefont {van
  Leeuwen}}, \ and\ \bibinfo {author} {\bibfnamefont {E.~K.~U.}\ \bibnamefont
  {Gross}},\ }\href@noop {} {\bibfield  {journal} {\bibinfo  {journal} {Phys.
  Rev. A}\ }\textbf {\bibinfo {volume} {78}},\ \bibinfo {pages} {022501}
  (\bibinfo {year} {2008})}\BibitemShut {NoStop}%
\bibitem [{\citenamefont {Kreibich}\ \emph {et~al.}(2004)\citenamefont
  {Kreibich}, \citenamefont {van Leeuwen},\ and\ \citenamefont {Gross}}]{en1}%
  \BibitemOpen
  \bibfield  {author} {\bibinfo {author} {\bibfnamefont {T.}~\bibnamefont
  {Kreibich}}, \bibinfo {author} {\bibfnamefont {R.}~\bibnamefont {van
  Leeuwen}}, \ and\ \bibinfo {author} {\bibfnamefont {E.~K.~U.}\ \bibnamefont
  {Gross}},\ }\href@noop {} {\bibfield  {journal} {\bibinfo  {journal} {Chem.
  Phys.}\ }\textbf {\bibinfo {volume} {304}},\ \bibinfo {pages} {183} (\bibinfo
  {year} {2004})}\BibitemShut {NoStop}%
\bibitem [{\citenamefont {Chakraborty}\ \emph {et~al.}(2008)\citenamefont
  {Chakraborty}, \citenamefont {Pak},\ and\ \citenamefont
  {Hammes-Schiffer}}]{en2}%
  \BibitemOpen
  \bibfield  {author} {\bibinfo {author} {\bibfnamefont {A.}~\bibnamefont
  {Chakraborty}}, \bibinfo {author} {\bibfnamefont {M.~V.}\ \bibnamefont
  {Pak}}, \ and\ \bibinfo {author} {\bibfnamefont {S.}~\bibnamefont
  {Hammes-Schiffer}},\ }\href@noop {} {\bibfield  {journal} {\bibinfo
  {journal} {Phys. Rev. Lett.}\ }\textbf {\bibinfo {volume} {101}},\ \bibinfo
  {pages} {153001} (\bibinfo {year} {2008})}\BibitemShut {NoStop}%
\bibitem [{\citenamefont {Udagawa}\ \emph {et~al.}(2014)\citenamefont
  {Udagawa}, \citenamefont {Tsuneda},\ and\ \citenamefont {Tachikawa}}]{en3}%
  \BibitemOpen
  \bibfield  {author} {\bibinfo {author} {\bibfnamefont {T.}~\bibnamefont
  {Udagawa}}, \bibinfo {author} {\bibfnamefont {T.}~\bibnamefont {Tsuneda}}, \
  and\ \bibinfo {author} {\bibfnamefont {M.}~\bibnamefont {Tachikawa}},\
  }\href@noop {} {\bibfield  {journal} {\bibinfo  {journal} {Phys. Rev. A}\
  }\textbf {\bibinfo {volume} {89}},\ \bibinfo {pages} {052519} (\bibinfo
  {year} {2014})}\BibitemShut {NoStop}%
\bibitem [{\citenamefont {Abedi}\ \emph {et~al.}(2010)\citenamefont {Abedi},
  \citenamefont {Maitra},\ and\ \citenamefont {Gross}}]{ef1}%
  \BibitemOpen
  \bibfield  {author} {\bibinfo {author} {\bibfnamefont {A.}~\bibnamefont
  {Abedi}}, \bibinfo {author} {\bibfnamefont {N.~T.}\ \bibnamefont {Maitra}}, \
  and\ \bibinfo {author} {\bibfnamefont {E.~K.~U.}\ \bibnamefont {Gross}},\
  }\href@noop {} {\bibfield  {journal} {\bibinfo  {journal} {Phys. Rev. Lett.}\
  }\textbf {\bibinfo {volume} {105}},\ \bibinfo {pages} {123002} (\bibinfo
  {year} {2010})}\BibitemShut {NoStop}%
\bibitem [{\citenamefont {Suzuki}\ \emph {et~al.}(2014)\citenamefont {Suzuki},
  \citenamefont {Abedi}, \citenamefont {Maitra}, \citenamefont {Yamashita},\
  and\ \citenamefont {Gross}}]{ef2}%
  \BibitemOpen
  \bibfield  {author} {\bibinfo {author} {\bibfnamefont {Y.}~\bibnamefont
  {Suzuki}}, \bibinfo {author} {\bibfnamefont {A.}~\bibnamefont {Abedi}},
  \bibinfo {author} {\bibfnamefont {N.~T.}\ \bibnamefont {Maitra}}, \bibinfo
  {author} {\bibfnamefont {K.}~\bibnamefont {Yamashita}}, \ and\ \bibinfo
  {author} {\bibfnamefont {E.~K.~U.}\ \bibnamefont {Gross}},\ }\href@noop {}
  {\bibfield  {journal} {\bibinfo  {journal} {Phys. Rev. A}\ }\textbf {\bibinfo
  {volume} {89}},\ \bibinfo {pages} {040501} (\bibinfo {year}
  {2014})}\BibitemShut {NoStop}%
\bibitem [{\citenamefont {Bostr{\"o}m}\ \emph {et~al.}(2016)\citenamefont
  {Bostr{\"o}m}, \citenamefont {Mikkelsen},\ and\ \citenamefont
  {Verdozzi}}]{ef3}%
  \BibitemOpen
  \bibfield  {author} {\bibinfo {author} {\bibfnamefont {E.}~\bibnamefont
  {Bostr{\"o}m}}, \bibinfo {author} {\bibfnamefont {A.}~\bibnamefont
  {Mikkelsen}}, \ and\ \bibinfo {author} {\bibfnamefont {C.}~\bibnamefont
  {Verdozzi}},\ }\href@noop {} {\bibfield  {journal} {\bibinfo  {journal}
  {Phys. Rev. B}\ }\textbf {\bibinfo {volume} {93}},\ \bibinfo {pages} {195416}
  (\bibinfo {year} {2016})}\BibitemShut {NoStop}%
\bibitem [{\citenamefont {Requist}\ and\ \citenamefont {Gross}(2016)}]{ef4}%
  \BibitemOpen
  \bibfield  {author} {\bibinfo {author} {\bibfnamefont {R.}~\bibnamefont
  {Requist}}\ and\ \bibinfo {author} {\bibfnamefont {E.~K.~U.}\ \bibnamefont
  {Gross}},\ }\href@noop {} {\bibfield  {journal} {\bibinfo  {journal} {Phys.
  Rev. Lett.}\ }\textbf {\bibinfo {volume} {117}},\ \bibinfo {pages} {193001}
  (\bibinfo {year} {2016})}\BibitemShut {NoStop}%
\bibitem [{Note1()}]{Note1}%
  \BibitemOpen
  \bibinfo {note} {Note that it would also be possible to treat nuclei quantum
  mechanically by developing three-component density functional theory. In that
  case the body-fixed frame transformation would again need to be carried
  out}\BibitemShut {NoStop}%
\bibitem [{\citenamefont {Vignale}(2008)}]{action}%
  \BibitemOpen
  \bibfield  {author} {\bibinfo {author} {\bibfnamefont {G.}~\bibnamefont
  {Vignale}},\ }\href@noop {} {\bibfield  {journal} {\bibinfo  {journal} {Phys.
  Rev. A}\ }\textbf {\bibinfo {volume} {77}},\ \bibinfo {pages} {062511}
  (\bibinfo {year} {2008})}\BibitemShut {NoStop}%
\bibitem [{\citenamefont {van Leeuwen}(1998)}]{keldysh}%
  \BibitemOpen
  \bibfield  {author} {\bibinfo {author} {\bibfnamefont {R.}~\bibnamefont {van
  Leeuwen}},\ }\href@noop {} {\bibfield  {journal} {\bibinfo  {journal} {Phys.
  Rev. Lett.}\ }\textbf {\bibinfo {volume} {80}},\ \bibinfo {pages} {1280}
  (\bibinfo {year} {1998})}\BibitemShut {NoStop}%
\bibitem [{\citenamefont {Li}\ and\ \citenamefont
  {Ullrich}(2015)}]{aldastudy1}%
  \BibitemOpen
  \bibfield  {author} {\bibinfo {author} {\bibfnamefont {Y.}~\bibnamefont
  {Li}}\ and\ \bibinfo {author} {\bibfnamefont {C.~A.}\ \bibnamefont
  {Ullrich}},\ }\href@noop {} {\bibfield  {journal} {\bibinfo  {journal} {J.
  Chem. Theory Comput.}\ }\textbf {\bibinfo {volume} {11}},\ \bibinfo {pages}
  {5838} (\bibinfo {year} {2015})}\BibitemShut {NoStop}%
\bibitem [{\citenamefont {Dauth}\ \emph {et~al.}(2016)\citenamefont {Dauth},
  \citenamefont {Graus}, \citenamefont {Schelter}, \citenamefont {Wie{\ss}ner},
  \citenamefont {Sch{\"o}ll}, \citenamefont {Reinert},\ and\ \citenamefont
  {K{\"u}mmel}}]{aldastudy2}%
  \BibitemOpen
  \bibfield  {author} {\bibinfo {author} {\bibfnamefont {M.}~\bibnamefont
  {Dauth}}, \bibinfo {author} {\bibfnamefont {M.}~\bibnamefont {Graus}},
  \bibinfo {author} {\bibfnamefont {I.}~\bibnamefont {Schelter}}, \bibinfo
  {author} {\bibfnamefont {M.}~\bibnamefont {Wie{\ss}ner}}, \bibinfo {author}
  {\bibfnamefont {A.}~\bibnamefont {Sch{\"o}ll}}, \bibinfo {author}
  {\bibfnamefont {F.}~\bibnamefont {Reinert}}, \ and\ \bibinfo {author}
  {\bibfnamefont {S.}~\bibnamefont {K{\"u}mmel}},\ }\href@noop {} {\bibfield
  {journal} {\bibinfo  {journal} {Phys. Rev. Lett.}\ }\textbf {\bibinfo
  {volume} {117}},\ \bibinfo {pages} {183001} (\bibinfo {year}
  {2016})}\BibitemShut {NoStop}%
\bibitem [{\citenamefont {Giovanni}\ \emph {et~al.}(2017)\citenamefont
  {Giovanni}, \citenamefont {H{\"u}bener},\ and\ \citenamefont
  {Rubio}}]{aldastudy3}%
  \BibitemOpen
  \bibfield  {author} {\bibinfo {author} {\bibfnamefont {U.~D.}\ \bibnamefont
  {Giovanni}}, \bibinfo {author} {\bibfnamefont {H.}~\bibnamefont
  {H{\"u}bener}}, \ and\ \bibinfo {author} {\bibfnamefont {A.}~\bibnamefont
  {Rubio}},\ }\href@noop {} {\bibfield  {journal} {\bibinfo  {journal} {J.
  Chem. Theory Comput.}\ }\textbf {\bibinfo {volume} {13}},\ \bibinfo {pages}
  {265} (\bibinfo {year} {2017})}\BibitemShut {NoStop}%
\bibitem [{\citenamefont {Yabana}\ \emph {et~al.}(2012)\citenamefont {Yabana},
  \citenamefont {Sugiyama}, \citenamefont {Shinohara}, \citenamefont {Otobe},\
  and\ \citenamefont {Bertsch}}]{aldastudy4}%
  \BibitemOpen
  \bibfield  {author} {\bibinfo {author} {\bibfnamefont {K.}~\bibnamefont
  {Yabana}}, \bibinfo {author} {\bibfnamefont {T.}~\bibnamefont {Sugiyama}},
  \bibinfo {author} {\bibfnamefont {Y.}~\bibnamefont {Shinohara}}, \bibinfo
  {author} {\bibfnamefont {T.}~\bibnamefont {Otobe}}, \ and\ \bibinfo {author}
  {\bibfnamefont {G.~F.}\ \bibnamefont {Bertsch}},\ }\href@noop {} {\bibfield
  {journal} {\bibinfo  {journal} {Phys. Rev. B}\ }\textbf {\bibinfo {volume}
  {85}},\ \bibinfo {pages} {045134} (\bibinfo {year} {2012})}\BibitemShut
  {NoStop}%
\bibitem [{\citenamefont {Miyamoto}\ and\ \citenamefont
  {Rubio}(2018)}]{aldastudy5}%
  \BibitemOpen
  \bibfield  {author} {\bibinfo {author} {\bibfnamefont {Y.}~\bibnamefont
  {Miyamoto}}\ and\ \bibinfo {author} {\bibfnamefont {A.}~\bibnamefont
  {Rubio}},\ }\href@noop {} {\bibfield  {journal} {\bibinfo  {journal} {J.
  Phys. Soc. Jpn.}\ }\textbf {\bibinfo {volume} {87}},\ \bibinfo {pages}
  {041016} (\bibinfo {year} {2018})}\BibitemShut {NoStop}%
\bibitem [{\citenamefont {Ueda}\ \emph {et~al.}(2018)\citenamefont {Ueda},
  \citenamefont {Suzuki},\ and\ \citenamefont {Watanabe}}]{aldastudy6}%
  \BibitemOpen
  \bibfield  {author} {\bibinfo {author} {\bibfnamefont {Y.}~\bibnamefont
  {Ueda}}, \bibinfo {author} {\bibfnamefont {Y.}~\bibnamefont {Suzuki}}, \ and\
  \bibinfo {author} {\bibfnamefont {K.}~\bibnamefont {Watanabe}},\ }\href@noop
  {} {\bibfield  {journal} {\bibinfo  {journal} {Phys. Rev. B}\ }\textbf
  {\bibinfo {volume} {97}},\ \bibinfo {pages} {075406} (\bibinfo {year}
  {2018})}\BibitemShut {NoStop}%
\bibitem [{\citenamefont {Maitra}\ \emph {et~al.}(2002)\citenamefont {Maitra},
  \citenamefont {Burke},\ and\ \citenamefont {Woodward}}]{aldavalid1}%
  \BibitemOpen
  \bibfield  {author} {\bibinfo {author} {\bibfnamefont {N.~T.}\ \bibnamefont
  {Maitra}}, \bibinfo {author} {\bibfnamefont {K.}~\bibnamefont {Burke}}, \
  and\ \bibinfo {author} {\bibfnamefont {C.}~\bibnamefont {Woodward}},\
  }\href@noop {} {\bibfield  {journal} {\bibinfo  {journal} {Phys. Rev. Lett.}\
  }\textbf {\bibinfo {volume} {89}},\ \bibinfo {pages} {023002} (\bibinfo
  {year} {2002})}\BibitemShut {NoStop}%
\bibitem [{\citenamefont {Wijewardane}\ and\ \citenamefont
  {Ullrich}(2005)}]{aldavalid2}%
  \BibitemOpen
  \bibfield  {author} {\bibinfo {author} {\bibfnamefont {H.~O.}\ \bibnamefont
  {Wijewardane}}\ and\ \bibinfo {author} {\bibfnamefont {C.~A.}\ \bibnamefont
  {Ullrich}},\ }\href@noop {} {\bibfield  {journal} {\bibinfo  {journal} {Phys.
  Rev. Lett.}\ }\textbf {\bibinfo {volume} {95}},\ \bibinfo {pages} {086401}
  (\bibinfo {year} {2005})}\BibitemShut {NoStop}%
\bibitem [{\citenamefont {Thiele}\ \emph {et~al.}(2008)\citenamefont {Thiele},
  \citenamefont {Gross},\ and\ \citenamefont {K{\"u}mmel}}]{aldavalid3}%
  \BibitemOpen
  \bibfield  {author} {\bibinfo {author} {\bibfnamefont {M.}~\bibnamefont
  {Thiele}}, \bibinfo {author} {\bibfnamefont {E.~K.~U.}\ \bibnamefont
  {Gross}}, \ and\ \bibinfo {author} {\bibfnamefont {S.}~\bibnamefont
  {K{\"u}mmel}},\ }\href@noop {} {\bibfield  {journal} {\bibinfo  {journal}
  {Phys. Rev. Lett.}\ }\textbf {\bibinfo {volume} {100}},\ \bibinfo {pages}
  {153004} (\bibinfo {year} {2008})}\BibitemShut {NoStop}%
\bibitem [{\citenamefont {Hofmann}\ \emph {et~al.}(2012)\citenamefont
  {Hofmann}, \citenamefont {K{\"o}rzd{\"o}rfer},\ and\ \citenamefont
  {K{\"u}mmel}}]{aldavalid4}%
  \BibitemOpen
  \bibfield  {author} {\bibinfo {author} {\bibfnamefont {D.}~\bibnamefont
  {Hofmann}}, \bibinfo {author} {\bibfnamefont {T.}~\bibnamefont
  {K{\"o}rzd{\"o}rfer}}, \ and\ \bibinfo {author} {\bibfnamefont
  {S.}~\bibnamefont {K{\"u}mmel}},\ }\href@noop {} {\bibfield  {journal}
  {\bibinfo  {journal} {Phys. Rev. Lett.}\ }\textbf {\bibinfo {volume} {108}},\
  \bibinfo {pages} {146401} (\bibinfo {year} {2012})}\BibitemShut {NoStop}%
\bibitem [{\citenamefont {Ramsden}\ and\ \citenamefont
  {Godby}(2012)}]{aldavalid5}%
  \BibitemOpen
  \bibfield  {author} {\bibinfo {author} {\bibfnamefont {J.~D.}\ \bibnamefont
  {Ramsden}}\ and\ \bibinfo {author} {\bibfnamefont {R.~W.}\ \bibnamefont
  {Godby}},\ }\href@noop {} {\bibfield  {journal} {\bibinfo  {journal} {Phys.
  Rev. Lett.}\ }\textbf {\bibinfo {volume} {109}},\ \bibinfo {pages} {036402}
  (\bibinfo {year} {2012})}\BibitemShut {NoStop}%
\bibitem [{\citenamefont {Helbig}\ \emph {et~al.}(2011)\citenamefont {Helbig},
  \citenamefont {Fuks}, \citenamefont {Casula}, \citenamefont {Verstraete},
  \citenamefont {Marques}, \citenamefont {Tokatly},\ and\ \citenamefont
  {Rubio}}]{aldavalid6}%
  \BibitemOpen
  \bibfield  {author} {\bibinfo {author} {\bibfnamefont {N.}~\bibnamefont
  {Helbig}}, \bibinfo {author} {\bibfnamefont {J.~I.}\ \bibnamefont {Fuks}},
  \bibinfo {author} {\bibfnamefont {M.}~\bibnamefont {Casula}}, \bibinfo
  {author} {\bibfnamefont {M.~J.}\ \bibnamefont {Verstraete}}, \bibinfo
  {author} {\bibfnamefont {M.~A.~L.}\ \bibnamefont {Marques}}, \bibinfo
  {author} {\bibfnamefont {I.~V.}\ \bibnamefont {Tokatly}}, \ and\ \bibinfo
  {author} {\bibfnamefont {A.}~\bibnamefont {Rubio}},\ }\href@noop {}
  {\bibfield  {journal} {\bibinfo  {journal} {Phys. Rev. A}\ }\textbf {\bibinfo
  {volume} {83}},\ \bibinfo {pages} {032503} (\bibinfo {year}
  {2011})}\BibitemShut {NoStop}%
\bibitem [{\citenamefont {Suzuki}\ \emph {et~al.}(2017)\citenamefont {Suzuki},
  \citenamefont {Lacombe}, \citenamefont {Watanabe},\ and\ \citenamefont
  {Maitra}}]{aldavalid7}%
  \BibitemOpen
  \bibfield  {author} {\bibinfo {author} {\bibfnamefont {Y.}~\bibnamefont
  {Suzuki}}, \bibinfo {author} {\bibfnamefont {L.}~\bibnamefont {Lacombe}},
  \bibinfo {author} {\bibfnamefont {K.}~\bibnamefont {Watanabe}}, \ and\
  \bibinfo {author} {\bibfnamefont {N.~T.}\ \bibnamefont {Maitra}},\
  }\href@noop {} {\bibfield  {journal} {\bibinfo  {journal} {Phys. Rev. Lett.}\
  }\textbf {\bibinfo {volume} {119}},\ \bibinfo {pages} {263401} (\bibinfo
  {year} {2017})}\BibitemShut {NoStop}%
\bibitem [{\citenamefont {Perdew}\ and\ \citenamefont {Zunger}(1981)}]{lda}%
  \BibitemOpen
  \bibfield  {author} {\bibinfo {author} {\bibfnamefont {J.~P.}\ \bibnamefont
  {Perdew}}\ and\ \bibinfo {author} {\bibfnamefont {A.}~\bibnamefont
  {Zunger}},\ }\href@noop {} {\bibfield  {journal} {\bibinfo  {journal} {Phys.
  Rev. B}\ }\textbf {\bibinfo {volume} {23}},\ \bibinfo {pages} {5048}
  (\bibinfo {year} {1981})}\BibitemShut {NoStop}%
\bibitem [{\citenamefont {Gianturco}\ \emph {et~al.}(2006)\citenamefont
  {Gianturco}, \citenamefont {Franz}, \citenamefont {Buenker}, \citenamefont
  {Liebermann}, \citenamefont {Pichl}, \citenamefont {Rost}, \citenamefont
  {Tachikawa},\ and\ \citenamefont {Kimura}}]{lih1}%
  \BibitemOpen
  \bibfield  {author} {\bibinfo {author} {\bibfnamefont {F.~A.}\ \bibnamefont
  {Gianturco}}, \bibinfo {author} {\bibfnamefont {J.}~\bibnamefont {Franz}},
  \bibinfo {author} {\bibfnamefont {R.~J.}\ \bibnamefont {Buenker}}, \bibinfo
  {author} {\bibfnamefont {H.-P.}\ \bibnamefont {Liebermann}}, \bibinfo
  {author} {\bibfnamefont {L.}~\bibnamefont {Pichl}}, \bibinfo {author}
  {\bibfnamefont {J.-M.}\ \bibnamefont {Rost}}, \bibinfo {author}
  {\bibfnamefont {M.}~\bibnamefont {Tachikawa}}, \ and\ \bibinfo {author}
  {\bibfnamefont {M.}~\bibnamefont {Kimura}},\ }\href@noop {} {\bibfield
  {journal} {\bibinfo  {journal} {Phys. Rev. A}\ }\textbf {\bibinfo {volume}
  {73}},\ \bibinfo {pages} {022705} (\bibinfo {year} {2006})}\BibitemShut
  {NoStop}%
\bibitem [{\citenamefont {Kita}\ \emph {et~al.}(2011)\citenamefont {Kita},
  \citenamefont {Maezono}, \citenamefont {Tachikawa}, \citenamefont {Toeler},\
  and\ \citenamefont {Needs}}]{lih2}%
  \BibitemOpen
  \bibfield  {author} {\bibinfo {author} {\bibfnamefont {Y.}~\bibnamefont
  {Kita}}, \bibinfo {author} {\bibfnamefont {R.}~\bibnamefont {Maezono}},
  \bibinfo {author} {\bibfnamefont {M.}~\bibnamefont {Tachikawa}}, \bibinfo
  {author} {\bibfnamefont {M.~D.}\ \bibnamefont {Toeler}}, \ and\ \bibinfo
  {author} {\bibfnamefont {R.~J.}\ \bibnamefont {Needs}},\ }\href@noop {}
  {\bibfield  {journal} {\bibinfo  {journal} {J. Chem. Phys.}\ }\textbf
  {\bibinfo {volume} {135}},\ \bibinfo {pages} {054108} (\bibinfo {year}
  {2011})}\BibitemShut {NoStop}%
\bibitem [{\citenamefont {Sirjoosingh}\ \emph {et~al.}(2013)\citenamefont
  {Sirjoosingh}, \citenamefont {Pak}, \citenamefont {Swalina},\ and\
  \citenamefont {Hammes-Schiffer}}]{lih3}%
  \BibitemOpen
  \bibfield  {author} {\bibinfo {author} {\bibfnamefont {A.}~\bibnamefont
  {Sirjoosingh}}, \bibinfo {author} {\bibfnamefont {M.~V.}\ \bibnamefont
  {Pak}}, \bibinfo {author} {\bibfnamefont {C.}~\bibnamefont {Swalina}}, \ and\
  \bibinfo {author} {\bibfnamefont {S.}~\bibnamefont {Hammes-Schiffer}},\
  }\href@noop {} {\bibfield  {journal} {\bibinfo  {journal} {J. Chem. Phys.}\
  }\textbf {\bibinfo {volume} {139}},\ \bibinfo {pages} {034103} (\bibinfo
  {year} {2013})}\BibitemShut {NoStop}%
\bibitem [{dis(2018)}]{distance}%
  \BibitemOpen
  \href@noop {} {\emph {\bibinfo {title} {NIST Computational Chemistry
  Comparison and Benchmark Database}}} (\bibinfo {year} {2018}),\ \bibinfo
  {note} {\url{https://cccbdb.nist.gov/}}\BibitemShut {NoStop}%
\bibitem [{\citenamefont {Kobayashi}(1999)}]{ncpp}%
  \BibitemOpen
  \bibfield  {author} {\bibinfo {author} {\bibfnamefont {K.}~\bibnamefont
  {Kobayashi}},\ }\href@noop {} {\bibfield  {journal} {\bibinfo  {journal}
  {Comput. Mater. Sci.}\ }\textbf {\bibinfo {volume} {14}},\ \bibinfo {pages}
  {72} (\bibinfo {year} {1999})}\BibitemShut {NoStop}%
\bibitem [{Note2()}]{Note2}%
  \BibitemOpen
  \bibinfo {note} {Movies of the dynamics are given in the Supplemental
  Material.}\BibitemShut {Stop}%
\bibitem [{Note3()}]{Note3}%
  \BibitemOpen
  \bibinfo {note} {Note that steps appeared in $v_{\protect \rm KS}^-$ around
  $x=5-10$ \r Aare due to the LDA electron-positron correlation energy
  functional parameterized in Ref.~\cite {tcdftatom}}\BibitemShut {NoStop}%
\bibitem [{Note4()}]{Note4}%
  \BibitemOpen
  \bibinfo {note} {The bound region is defined as the region where considerable
  ground-state positron density exists (the region where $n^+(x,
  t=0)>0.001$)}\BibitemShut {NoStop}%
\bibitem [{Note5()}]{Note5}%
  \BibitemOpen
  \bibinfo {note} {Time-dependent dipole moment of LiH without positronic
  contribution $d_x^-(t)$ is given by $d_x^-(t)=-\DOTSI \intop \ilimits@
  d{\protect \bf r}^- x n^-({\protect \bf r}^-, t)$}\BibitemShut {NoStop}%
\bibitem [{\citenamefont {Casida}(1995)}]{casida}%
  \BibitemOpen
  \bibfield  {author} {\bibinfo {author} {\bibfnamefont {M.~E.}\ \bibnamefont
  {Casida}},\ }in\ \href@noop {} {\emph {\bibinfo {booktitle} {Recent
  {Advances} in {Density} {Functional} {Methods}}}}\ (\bibinfo  {publisher}
  {World Scientific, Singapore},\ \bibinfo {year} {1995})\ pp.\ \bibinfo
  {pages} {155--192}\BibitemShut {NoStop}%
\bibitem [{\citenamefont {Ullrich}\ \emph {et~al.}(1997)\citenamefont
  {Ullrich}, \citenamefont {Reinhard},\ and\ \citenamefont
  {Suraud}}]{resonant1}%
  \BibitemOpen
  \bibfield  {author} {\bibinfo {author} {\bibfnamefont {C.~A.}\ \bibnamefont
  {Ullrich}}, \bibinfo {author} {\bibfnamefont {P.-G.}\ \bibnamefont
  {Reinhard}}, \ and\ \bibinfo {author} {\bibfnamefont {E.}~\bibnamefont
  {Suraud}},\ }\href@noop {} {\bibfield  {journal} {\bibinfo  {journal} {J.
  Phys. B}\ }\textbf {\bibinfo {volume} {30}},\ \bibinfo {pages} {5043}
  (\bibinfo {year} {1997})}\BibitemShut {NoStop}%
\bibitem [{\citenamefont {Ma}\ \emph {et~al.}(2015)\citenamefont {Ma},
  \citenamefont {Wang},\ and\ \citenamefont {Wang}}]{resonant2}%
  \BibitemOpen
  \bibfield  {author} {\bibinfo {author} {\bibfnamefont {J.}~\bibnamefont
  {Ma}}, \bibinfo {author} {\bibfnamefont {Z.}~\bibnamefont {Wang}}, \ and\
  \bibinfo {author} {\bibfnamefont {L.-W.}\ \bibnamefont {Wang}},\ }\href@noop
  {} {\bibfield  {journal} {\bibinfo  {journal} {Nat. Commn.}\ }\textbf
  {\bibinfo {volume} {6}},\ \bibinfo {pages} {10107} (\bibinfo {year}
  {2015})}\BibitemShut {NoStop}%
\bibitem [{\citenamefont {Fuks}\ \emph {et~al.}(2011)\citenamefont {Fuks},
  \citenamefont {Rubio},\ and\ \citenamefont {Maitra}}]{ct1}%
  \BibitemOpen
  \bibfield  {author} {\bibinfo {author} {\bibfnamefont {J.~I.}\ \bibnamefont
  {Fuks}}, \bibinfo {author} {\bibfnamefont {A.}~\bibnamefont {Rubio}}, \ and\
  \bibinfo {author} {\bibfnamefont {N.~T.}\ \bibnamefont {Maitra}},\
  }\href@noop {} {\bibfield  {journal} {\bibinfo  {journal} {Phys. Rev. A}\
  }\textbf {\bibinfo {volume} {83}},\ \bibinfo {pages} {042501} (\bibinfo
  {year} {2011})}\BibitemShut {NoStop}%
\bibitem [{\citenamefont {Maitra}(2017)}]{ct2}%
  \BibitemOpen
  \bibfield  {author} {\bibinfo {author} {\bibfnamefont {N.~T.}\ \bibnamefont
  {Maitra}},\ }\href@noop {} {\bibfield  {journal} {\bibinfo  {journal} {J.
  Phys.: Condens. Matter}\ }\textbf {\bibinfo {volume} {29}},\ \bibinfo {pages}
  {423001} (\bibinfo {year} {2017})}\BibitemShut {NoStop}%
\end{thebibliography}%

\end{document}